\RequirePackage{rotating}
\documentclass[twocolumn, tighten]{aastex62}
\usepackage{amsmath}
\usepackage[caption=false]{subfig}
\usepackage{xfrac}

\shorttitle{ALMA Observations of GM Aur}
\shortauthors{Huang et al.}

\begin{document}

\title{ A multi-frequency ALMA characterization of substructures in the GM Aur protoplanetary disk }

\correspondingauthor{Jane Huang}
\email{jane.huang@cfa.harvard.edu}

\author{Jane Huang}
\affiliation{Center for Astrophysics $\mid$ Harvard \& Smithsonian, 60 Garden Street, Cambridge, MA 02138, United States of America}

\author{Sean M. Andrews}
\affiliation{Center for Astrophysics $\mid$ Harvard \& Smithsonian, 60 Garden Street, Cambridge, MA 02138, United States of America}

\author{Cornelis P. Dullemond}
\affiliation{Zentrum f\"ur Astronomie, Heidelberg University, Albert Ueberle Str. 2, 69120 Heidelberg, Germany}

\author{Karin I. \"Oberg}
\affiliation{Center for Astrophysics $\mid$ Harvard \& Smithsonian, 60 Garden Street, Cambridge, MA 02138, United States of America}

\author{Chunhua Qi}
\affiliation{Center for Astrophysics $\mid$ Harvard \& Smithsonian, 60 Garden Street, Cambridge, MA 02138, United States of America}

\author{Zhaohuan Zhu}
\affiliation{Department of Physics and Astronomy, University of Nevada, Las Vegas, 4505 S. Maryland Pkwy, Las Vegas, NV, 89154, USA}

\author{Tilman Birnstiel}
\affiliation{University Observatory, Faculty of Physics, Ludwig-Maximilians-Universit\"at M\"unchen, Scheinerstr. 1, 81679 Munich, Germany}

\author{John M. Carpenter}
\affiliation{Joint ALMA Observatory, Avenida Alonso de C\'ordova 3107, Vitacura, Santiago, Chile}

\author{Andrea Isella}
\affiliation{Department of Physics and Astronomy, Rice University, 6100 Main Street, Houston, TX 77005, United States of America}

\author{Enrique Mac\'ias}
\affiliation{Joint ALMA Observatory, Avenida Alonso de C\'ordova 3107, Vitacura, Santiago, Chile}
\affiliation{European Southern Observatory, Alonso de C\'ordova 3107, Vitacura, Santiago 763-0355, Chile}

\author{Melissa K. McClure}
\affiliation{Leiden Observatory, Leiden University, P.O. Box 9513, 2300 RA, Leiden, The Netherlands}

\author{Laura M. P\'erez}
\affiliation{Departamento de Astronom\'ia, Universidad de Chile, Camino El Observatorio 1515, Las Condes, Santiago, Chile}

\author{Richard Teague}
\affiliation{Center for Astrophysics $\mid$ Harvard \& Smithsonian, 60 Garden Street, Cambridge, MA 02138, United States of America}

\author{David J. Wilner}
\affiliation{Center for Astrophysics $\mid$ Harvard \& Smithsonian, 60 Garden Street, Cambridge, MA 02138, United States of America}

\author{Shangjia Zhang}
\affiliation{Department of Physics and Astronomy, University of Nevada, Las Vegas, 4505 S. Maryland Pkwy, Las Vegas, NV, 89154, USA}

\begin{abstract}
The protoplanetary disk around the T Tauri star GM Aur was one of the first hypothesized to be in the midst of being cleared out by a forming planet. As a result, GM Aur has had an outsized influence on our understanding of disk structure and evolution. We present 1.1 and 2.1 mm ALMA continuum observations of the GM Aur disk at a resolution of $\sim50$ mas ($\sim8$ au), as well as HCO$^+$ $J=3-2$ observations at a resolution of $\sim100$ mas. The dust continuum shows at least three rings atop faint, extended emission. Unresolved emission is detected at the center of the disk cavity at both wavelengths, likely due to a combination of dust and free-free emission. Compared to the 1.1 mm image, the 2.1 mm image shows a more pronounced ``shoulder'' near $R\sim40$ au, highlighting the utility of longer-wavelength observations for characterizing disk substructures. The spectral index $\alpha$ features strong radial variations, with minima near the emission peaks and maxima near the gaps. While low spectral indices have often been ascribed to grain growth and dust trapping, the optical depth of GM Aur's inner two emission rings renders their dust properties ambiguous. The gaps and outer disk ($R>100$ au) are optically thin at both wavelengths.  Meanwhile, the HCO$^+$ emission indicates that the gas cavity is more compact than the dust cavity traced by the millimeter continuum, similar to other disks traditionally classified as ``transitional.'' 
\end{abstract}


\section{Introduction} \label{sec:intro}

The dust, gas, and ice in the disks around young stars serve as the starting material for planet formation. Analyses of the spectral energy distributions (SEDs) of pre-main sequence stars provided early insights into disk evolution. Surveys indicated that low-mass pre-main sequence stars younger than a few Myr were typically surrounded by disks emitting brightly at infrared wavelengths, while stars older than 10 Myr seldomly appeared to have disks \citep[e.g.,][]{1989AJ.....97.1451S}. About 10\% of disks exhibited weak near-IR emission in conjunction with strong mid-IR and far-IR emission, which was interpreted as a signature of a brief ($\sim0.3$ Myr) ``transitional'' period in which disks developed a cavity before being fully dispersed from the inside-out \citep[e.g.,][]{1989AJ.....97.1451S, 1990AJ.....99.1187S}. Meanwhile, single-dish millimeter wavelength spectral index measurements of disks yielded values lower than that of the interstellar medium, suggesting that coagulation processes increased dust grain sizes in disks \citep[e.g.,][]{1990AJ.....99..924B}.

While spatially unresolved observations were the foundation of early disk characterizations, the improving resolution and sensitivity of millimeter interferometers have enabled the evolution of gas and dust in disks to be probed in unprecedented detail. Line observations mapped the Keplerian rotation of gas in disks and demonstrated that some disks stretched out as far as 1000 au \citep[e.g.][]{1991ApJ...382L..31S, 1993Icar..106....2K}. The cavities inferred from ``transition'' disk SEDS were resolved for the first time at millimeter wavelengths \citep[e.g.,][]{2008ApJ...675L.109B, 2011ApJ...732...42A}. Interferometric spectral index measurements suggested grain growth up to centimeter scales in disks \citep[e.g.,][]{2003AA...403..323T}. Initial detections of radial spectral index variations were interpreted to be the result of larger dust grains drifting toward the star more quickly (i.e., radial drift), a process that limits the timescale for solids to grow into planets \citep[e.g.,][]{2011AA...529A.105G, 2012ApJ...760L..17P}. 

\begin{deluxetable*}{ccccccc}
\tablecaption{ALMA Observing Summary \label{tab:observations}}
\tablehead{\colhead{Program ID}&\colhead{Date} &\colhead{Configuration}&\colhead{Freq. range} &\colhead{Antennas} & \colhead{Baselines}&\colhead{Time on source}\\
&&&\colhead{(GHz)}&&\colhead{(m)}&\colhead{(min)}}
\colnumbers
\startdata
\hline
\multicolumn{7}{c}{Band 4 Observations} \\
\hline
2017.1.01151.S& 2017 October 27 \tablenotemark{a}&Extended&$136.995-152.995$&47&$135-14900$&64\\
2017.1.01151.S& 2018 October 4&Compact&$136.995-152.995$&48&$15-2517$&32\\
\hline
\multicolumn{7}{c}{Band 6 Observations} \\
\hline
2017.1.01151.S& 2017 October 25 &Extended&$251.505-270.005$&44&$41-14900$&47\\
2017.1.01151.S& 2017 October 28&Extended&$251.504-270.004$&49&$113-13900$&47\\
2018.1.01230.S & 2018 October 18& Compact & $251.507-270.007$&47&$15-2517$&28
\enddata
\tablenotetext{a}{Two execution blocks were taken on this date}
\end{deluxetable*}

The advent of high resolution imaging by ALMA revealed that many disks, not just the $\sim$10\% classified as ``transitional,'' have their dust (and sometimes gas) arranged into complex structures \citep[e.g.,][]{2015ApJ...808L...3A, 2018ApJ...869L..41A}. The most commonly observed structures are annular gaps and rings \citep[e.g.,][]{2018ApJ...869...17L,2018ApJ...869L..42H}. For the few disks observed at widely-separated frequencies at high resolution (i.e., better than 10 au), it became apparent that radial spectral index variations  were intimately linked with the annular dust substructures, perhaps due either to the concentration of larger dust grains within gas pressure bumps \citep[e.g.,][]{2016ApJ...829L..35T,2019ApJ...881..159M} or to optical depth variations \citep[e.g.,][]{2018ApJ...852..122H,2019ApJ...877L..22L}. Distinguishing between these scenarios is necessary to clarify where the growth of solids occurs and how much solid material is available for planet formation. Increasing the sample of disks with multi-frequency high resolution observations is essential for determining if and how spectral indices vary with disk and stellar properties, which can in turn yield insights into what mechanism sets the spectral indices. 

One of the most well-studied pre-main sequence stars is GM Aur (ICRS 04$^\text{h}$55$^\text{m}$10$^\text{s}$.981, 30$^\circ$21$'$59$\farcs$376), a {$2.5\substack{+1.5\\-0.9}$ Myr old K6 star that is located $159\pm4$ pc away in the Taurus-Auriga star-forming region and hosts a large protoplanetary disk \citep[e.g.,][]{1993Icar..106....2K, 2014ApJ...786...97H, 2018AA...616A...1G, 2018ApJ...865..157A}. GM Aur was among the minority of T Tauri stars that \citet{1989AJ.....97.1451S} noted for their weak near-IR emission relative to typical T Tauri stars, which motivated the introduction of the ``transition disk'' concept. Based on SED modeling, \citet{1992ApJ...395L.115M} hypothesized that GM Aur's disk cavity was opened by one or more planets. Based on apparent deviations from Keplerian rotation in low resolution CO observations, \citet{2008AA...490L..15D} and \citet{2009ApJ...698..131H} hypothesized that the GM Aur disk is warped by a planet. Millimeter interferometric observations  have resolved a dust cavity with a radius of $\sim40$ au and revealed additional annular dust substructures \citep[e.g.,][]{2009ApJ...698..131H, 2018ApJ...865...37M}.

In this work, we present new high resolution ALMA continuum observations of the GM Aur disk at 2.1 and 1.1 mm in conjunction with HCO$^+$ $J=3-2$ observations. The continuum observations reveal new dust substructures and are used to constrain the dust grain properties. HCO$^+$ was targeted to investigate previous claims of a significant kinematic disturbance in the GM Aur disk with a line that is not afflicted by cloud contamination. Section \ref{sec:observations} describes the observations and data reduction. An overview of the continuum emission is presented in Section \ref{sec:results}. Modeling and analysis of the continuum substructures and dust properties are presented in Section \ref{sec:continuummodels}. Analysis of the HCO$^+$ emission is presented in Section \ref{sec:hcop}. Our results are discussed in Section \ref{sec:discussion}. Section \ref{sec:summary} summarizes the main findings.

\begin{deluxetable*}{ccccccc}
\tablecaption{Imaging summary\label{tab:imageproperties}}
\tablehead{
&\colhead{Frequency}&\colhead{Briggs parameter}&\colhead{Synthesized beam}&\colhead{Peak $I_\nu$}&\colhead{RMS noise}&\colhead{Flux\tablenotemark{a}}\\
&(GHz)&&(mas $\times$ mas ($^\circ$))&(mJy beam$^{-1}$)&(mJy beam$^{-1}$)&\colhead{(mJy)}}\startdata
2.1 mm continuum (B4) & 144.988 & 0 & $57\times 34$ $(-13.2)$ &0.49&0.012&$54.8\pm0.7$ \\
1.1 mm continuum (B6) & 260.745 & 0.5 & $45\times25$ $(2.2)$&0.94 & 0.01 &$264.1\pm0.8$\\
HCO$^+$ $J=3-2$ \tablenotemark{b} & 267.5576259\tablenotemark{c} & 1.0 & $107\times83$ $(6.4)$ &80 & 1.2& $6960\pm60$\\
\enddata
\tablenotetext{a}{Uncertainties do not include the $\sim10\%$ flux calibration uncertainty}
\tablenotetext{b} {For the HCO$^+$ data, peak $I_\nu$ (mJy beam$^{-1}$ km s$^{-1}$) and flux (mJy km s$^{-1}$) are reported for the integrated intensity map, while RMS noise (mJy beam$^{-1}$) is reported for the image cube with $dv=0.25$ km s$^{-1}$.} 
\tablenotetext{c}{From the Cologne Database for Molecular Spectroscopy \citep{2001AA...370L..49M, 2005JMoSt.742..215M}}
\end{deluxetable*}

\section{Observations and Data Reduction}\label{sec:observations}

ALMA observations of GM Aur were taken at 2.1 mm (Band 4) and 1.1 mm (Band 6), starting in Cycle 5 and completing as a Cycle 6 continuation program. The observation dates, antenna configuration properties, and time on-source are listed in Table \ref{tab:observations}. In each band, the disk was observed with an extended configuration to achieve high angular resolution and with a more compact configuration to recover larger scale emission. The Band 4 observations were set up with spectral windows (SPWs) centered at 137.995, 139.932, 149.995, and 151.995 GHz, each with a 2 GHz bandwidth and 15.625 MHz channel width. The extended configuration Band 6 observations from program 2017.1.01151.S were set up with SPWs centered at 252.505,  254.505, 267.655, and 269.005 GHz. All windows had bandwidths of 2 GHz and channel widths of 15.625 MHz, except for the window centered at 267.655 GHz, which had a bandwidth of 468.750 MHz and channel width of 122 kHz in order to spectrally resolve the HCO$^+$ $J=3-2$ line. The compact configuration Band 6 observation from program 2018.1.01230.S was set up with SPWs centered at 252.507, 254.507, 267.565, and 269.007 GHz. The window centered at 267.565 GHz had a bandwidth of 234.375 MHz and channel width of 61 kHz to target the HCO$^+$ $J=3-2$ line, while the other windows had bandwidths of 2 GHz and channel widths of 15.625 MHz. For all observations, the quasar J0510+1800 served as the bandpass and flux calibrator, while the quasar J0438+3004 served as the phase calibrator.

The raw data were calibrated with the ALMA pipeline. \texttt{CASA v.5.4.0} \citep{2007ASPC..376..127M} was used to perform additional data processing and imaging. The Band 4 data were time-averaged to 6-second intervals and frequency-averaged into 250 MHz-wide channels in order to reduce data volume. Since the compact and extended configuration observations were spatially offset from one another by a few hundredths of an arcsecond (as determined by comparing the centers of 2D Gaussians fits to the images in CASA), the \texttt{fixvis} and \texttt{fixplanets} tasks were used to align them by applying phase shifts and assigning common labels to the phase centers, respectively. The directions and magnitudes of the offsets are not consistent with a shift due purely to proper motion ($\mu_\alpha=3.899$, $\mu_\delta=-24.451$) \citep{2018AA...616A...1G}, so either atmospheric or instrumental effects are likely contributing to the small phase errors.  We checked that the visibility amplitudes from different dates were consistent within 5\% at overlapping spatial frequencies, which indicates that the fluxes are consistent between execution blocks. Phase and amplitude self-calibration were first applied to the compact configuration dataset alone using the multi-scale, multi-frequency synthesis imaging algorithm implemented in the  \texttt{tclean} task and scales of [$0''$, 0\farcs15, 0\farcs3, 0\farcs6, 0\farcs9]. An elliptical \texttt{CLEAN} mask with an orientation and aspect ratio similar to the continuum was used. The compact and extended data were then combined and phase self-calibrated together using scales of [$0''$, 0\farcs075, 0\farcs15, 0\farcs3, 0\farcs525]. We found that amplitude self-calibration did not improve the SNR of the high-resolution image. A Band 6 continuum image was produced in a similar manner, with the additional step beforehand of flagging channels covering the HCO$^+$ $J=3-2$ line. A lower Briggs parameter value (robust $=0$) was chosen for Band 4 imaging compared to the Band 6 imaging (robust $=0.5$) in order to achieve similar synthesized beams. 

The self-calibration solutions derived from the Band 6 continuum were applied to the HCO$^+$ $J=3-2$ SPWs. The \texttt{uvcontsub} task was used to subtract the continuum from the line emission in the $uv$ plane. An HCO$^+$ image cube with channel widths of 0.25 km s$^{-1}$ was produced using the \texttt{tclean} implementation of multi-scale with a Briggs robust value of 1.0 and a Gaussian outer taper to improve sensitivity to larger-scale emission. The selected CLEAN scales were [$0''$, 0\farcs2, 0\farcs5, 0\farcs75, 1\farcs5].  CLEAN masks were manually generated for individual channels to encompass the observed emission.

The continuum and HCO$^+$ image properties are summarized in Table \ref{tab:imageproperties}. The rms for each continuum image is measured inside an annulus that is centered on the disk and has an inner radius of $3''$ and outer radius of $5''$, which excludes all disk emission. The integrated flux is measured inside an elliptical mask with a position angle (P. A.) of $57\fdg17$, major axis of 2$''$, and a minor axis of $2''\times \cos i$, where $i=53\fdg21$. The P. A. and inclination $i$ are derived from the weighted average of fits to the 1.1 mm and 2.1 mm continuum profiles in the \textit{uv} plane (see Section \ref{sec:surfbrightness}). The mask major axis is selected through a method similar to that used in \citet{2016ApJ...828...46A}, where successively larger apertures are tested on the Band 6 image until the enclosed flux levels off. The flux uncertainty is computed with $\sqrt{\text{Area of mask}/\text{Area of beam}}\times \sigma$, where $\sigma$ is the image rms. The rms for the HCO$^+$ image is measured from line-free channels of the image cube. The procedure for measuring the HCO$^+$ integrated flux is described in more detail in Section \ref{sec:hcop}. 

The 1.1 mm (261 GHz) continuum flux measured with ALMA is within 10\% of the 267 GHz continuum flux measured with the Submillimeter Array in \citet{2010ApJ...720..480O}. These values are consistent given flux calibration uncertainties of $\sim10\%$ for each instrument, suggesting that the new ALMA observations adequately recover the flux. The 2.1 mm (145 GHz) continuum flux measured from the new ALMA data is $\sim45\%$ higher than the 141 GHz flux ($37\pm4$ mJy) measured with the Nobeyama Millimeter Array in \citet{2002ApJ...581..357K}. The shortest baselines from the Nobeyama observations are shorter than those of the ALMA observations, so the discrepancy cannot be attributed to spatial filtering of the Nobeyama data. However, the ALMA observations are much more sensitive, and the visibility amplitudes are consistent between the three execution blocks, suggesting that the flux calibration is reliable. Furthermore, the disk-averaged spectral index measured from the Bands 4 and 6 data are consistent with past measurements of GM Aur (see Section \ref{sec:results}).

\begin{figure*}
\begin{center}
\includegraphics[scale=0.95]{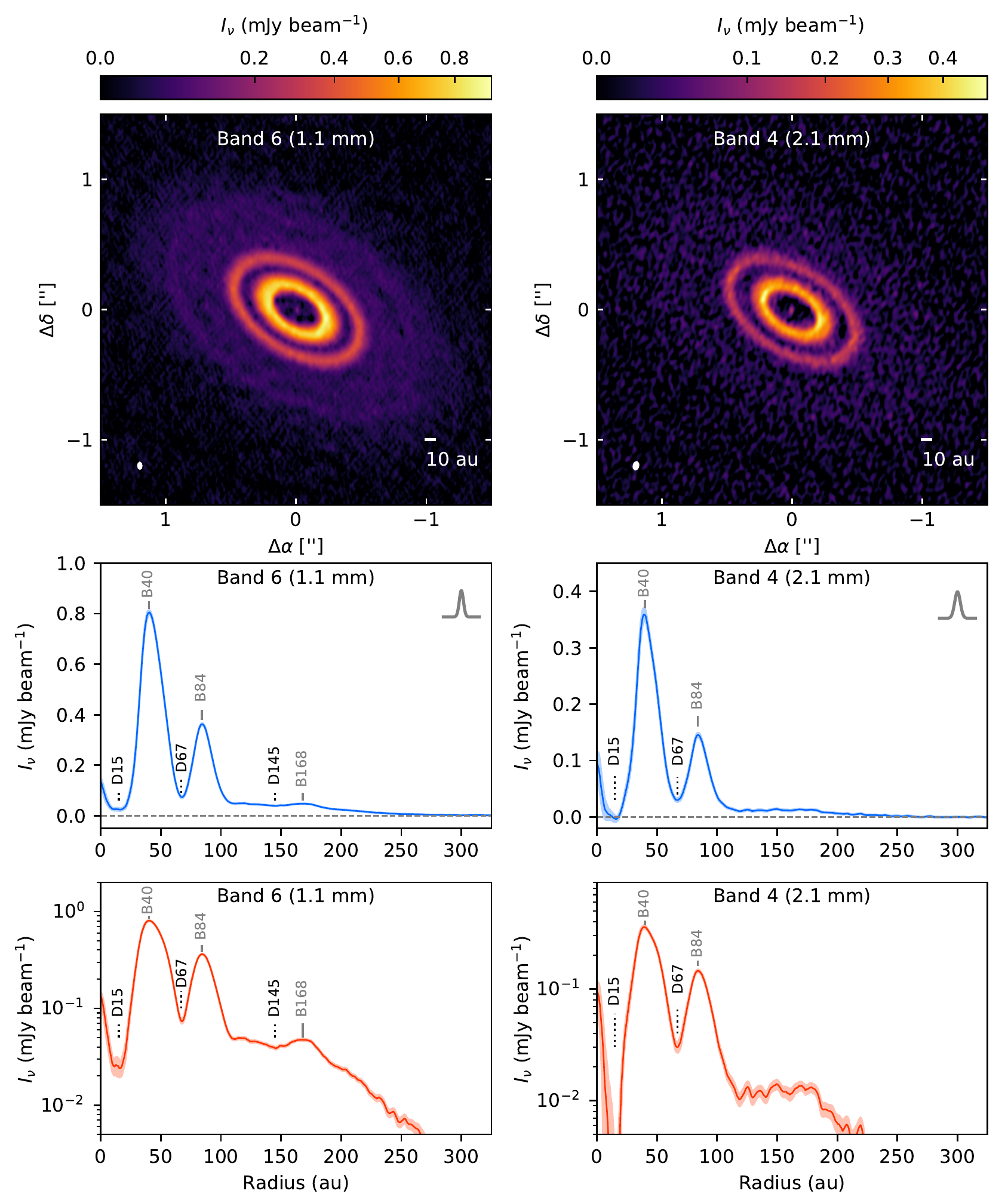}
\end{center}
\caption{\textit{Top row:} ALMA Band 6 (1.1 mm) and Band 4 (2.1 mm) continuum images of the GM Aur disk. A power-law normalization is used for the color scale to better display the faint outer disk emission. Axes show the angular offsets from the disk center. The synthesized beam is shown in the lower left corner of each image. \textit{Middle row:} Deprojected and azimuthally averaged radial intensity profiles for each band. Light blue shading shows the 1$\sigma$ scatter at each elliptical bin divided by the square root of the number of beams spanning the bin. The Gaussian profiles in the upper right corner of each panel show the width of the minor axis of the synthesized beams. \textit{Bottom row}: Radial intensity profiles replotted on a logarithmic scale. Light orange shading show the 1$\sigma$ scatter at each elliptical bin divided by the square root of the number of beams spanning the bin. The y-axis starts at $5\times10^{-3}$ mJy beam$^{-1}$, corresponding to slightly less than the rms of the continuum images. \label{fig:continuumoverview}}
\end{figure*}

\begin{figure*}
\begin{center}
\includegraphics{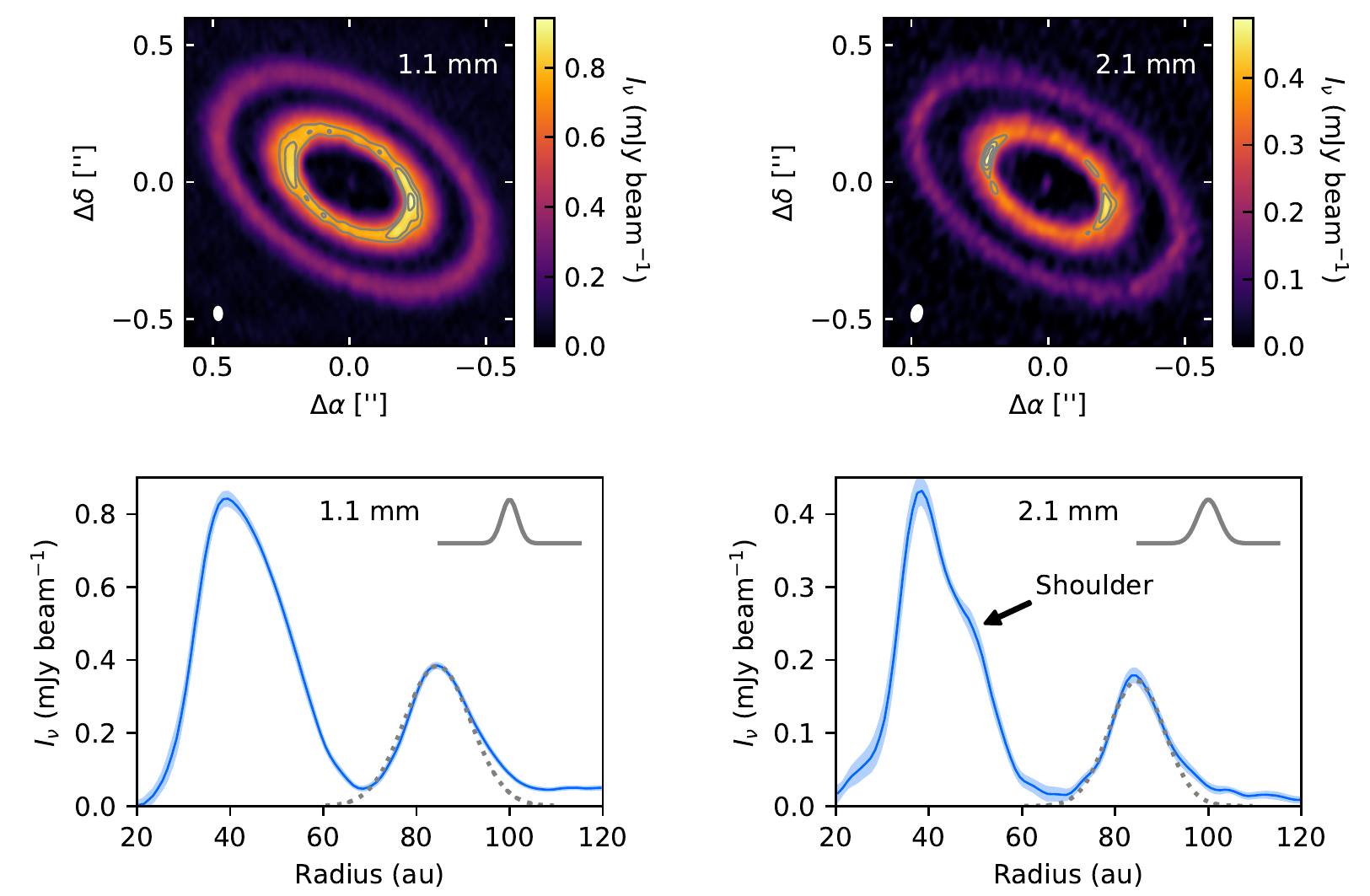}
\end{center}
\caption{\textit{Top row:} Inset of the 1.1 and 2.1 mm continuum images. Color bars are on a linear scale (unlike Figure \ref{fig:continuumoverview}, which uses power law scaling) to highlight the inner ring's structure. Contours are drawn at 75\%, 85\%, and 95\% of the peak intensity in each band, showing that B40 appears to be moderately brighter on the southwest side at 1.1 mm. Angular offsets from the disk center are marked on the axes. The synthesized beam is shown in the lower left corner of each image. \textit{Bottom row:} Deprojected radial profiles of the disk emission between 20 and 120 au, averaged across azimuthal angles extending $\pm20^\circ$ (in deprojected coordinates) from the projected major axis. Light blue shading shows the 1$\sigma$ scatter at each radial bin divided by the square root of the number of beams spanning the bin. The Gaussian profiles in the upper right corner of each panel show the width of the minor axis of the synthesized beams. At 1.1 mm, B40 is steeper on its interior side. At 2.1 mm, B40 has a ``shoulder'' on its outer edge that is not apparent at 1.1 mm. Best-fit Gaussian profiles are shown in dashed gray for B84 at both wavelengths to illustrate that the ring's inner edge is steeper compared to its outer edge. \label{fig:continuuminset}}
\end{figure*}

\section{Continuum emission properties \label{sec:results}}
\subsection{Continuum substructures}

The 1.1 and 2.1 mm ALMA continuum images, as well as their corresponding azimuthally averaged radial intensity profiles, are shown in Figure \ref{fig:continuumoverview}. The radial profiles are computed by deprojecting the continuum images using P. A. $=57\fdg17$ and $i=53\fdg21$, then averaging the pixel intensities within annular bins one au wide. The continuum observations in both bands reveal faint, compact emission at the center of the disk, bright narrow rings at $R\sim40$ and $\sim84$ au, and faint, diffuse emission beyond $R\sim100$ au. At 1.1 mm, a faint ring is visible at $R\sim168$ au on top of the outer diffuse emission. While the 2.1 mm image does not show an unambiguous counterpart to this ring, perhaps due to the lower signal-to-noise ratio (SNR) of the data, there appears to be a slow rise in emission toward $R\sim170$ au and a steeper falloff outside this radius. In accordance with the nomenclature from \citet{2018ApJ...869L..42H}, each ring is labeled with the prefix ``B'' (for ``bright'') followed by the radial location of the emission maximum rounded to the nearest whole number of astronomical units. The convention is similar for the gaps, except the prefix ``D'' (for ``dark'') is used. B40, D67, and B84, as well as the diffuse outer emission, were previously inferred from 930 $\mu$m observations at a resolution of $\sim0\farcs3$ in \citet{2018ApJ...865...37M}. D15 corresponds to the GM Aur disk's well-known central cavity \citep[e.g.,][]{2009ApJ...698..131H}, although the cavity might be more precisely described as an annular gap given the detection of interior emission. For the sake of continuity with previous works on GM Aur, we refer to this feature as the ``cavity'' in the rest of the paper. 

The new observations, which improve upon the resolution of previous GM Aur observations by an order of magnitude, show that B40 and B84 are not radially symmetric. This characteristic is more apparent in radial profiles generated from averaging only the pixels within 20$^\circ$ of the projected disk major axis (Figure \ref{fig:continuuminset}), especially since the synthesized beam is narrower along the disk major axis. At 1.1 mm, the emission profile of B40 is steeper on the side facing the star. At 2.1 mm, the emission profile of B40 appears to be narrower and more symmetric around the peak compared to the 1.1 mm image. Despite the slightly lower resolution of the 2.1 mm data, the appearance of an outer shoulder makes the two-component nature of B40 clearer. The differing emission profiles at the two wavelengths may either be due to the lower optical depth at 2.1 mm or to the 2.1 mm emission being more sensitive to larger grains, which may be confined to narrower regions by pressure bumps \citep{1972fpp..conf..211W}. The radial asymmetry of B84 is more subtle and is most easily highlighted by superimposing Gaussian profiles (derived by fitting the emission profiles between 80 and 90 au using the Levenberg-Marquardt minimization implementation in \texttt{scipy.optimize.curve\_fit}). In both bands, the B84 emission profile is shallower on the side facing away from the star. 

\begin{figure*}
\begin{center}
\includegraphics{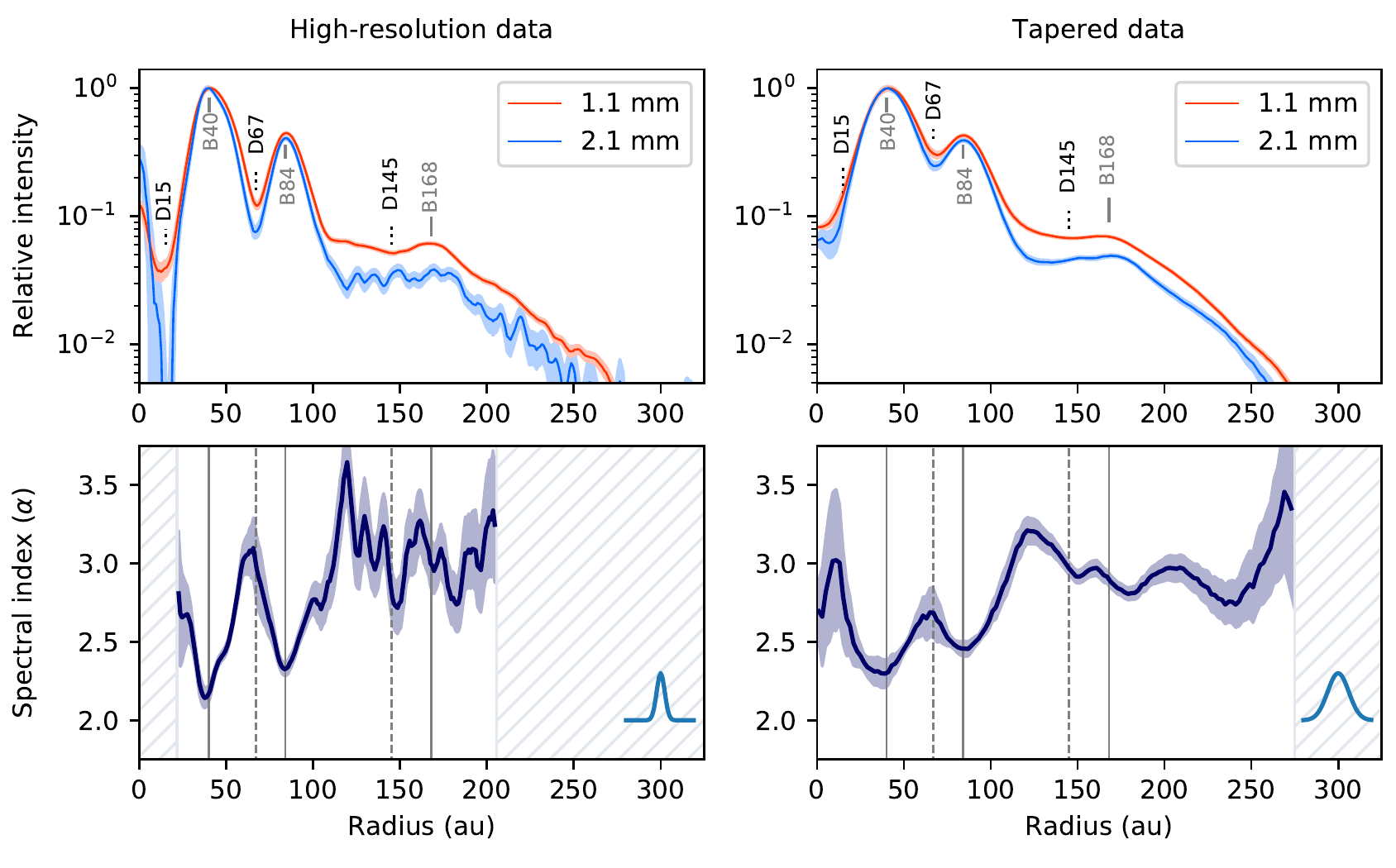}
\end{center}
\caption{\textit{Top left:} Azimuthally averaged radial intensity profiles of the high-resolution 2.1 and 1.1 mm continuum images, normalized to the peak of the respective radial profiles. The 1.1 mm profile is measured from an image smoothed to the same resolution as the 2.1 mm data (57 mas $\times$ 34 mas $(-13\fdg2)$). Shaded ribbons show the 1$\sigma$ scatter at each elliptical bin divided by the square root of the number of beams spanning the bin.  Solid gray lines mark the radial locations of the continuum rings, while dashed lines mark the continuum gaps. \textit{Bottom left:} Spectral index profiles measured from the above radial intensity profiles. Diagonally hatched boxes mark the regions where the spectral index cannot be estimated reliably from the image due to PSF artifacts in the inner disk or low SNR in the outer disk. The blue Gaussian profile shows the width of the minor axis of the synthesized beam. \textit{Top and bottom right:} Similar to plots on the left, except computed from images created by tapering and smoothing to a resolution of 110 mas $\times$ 90 mas.  \label{fig:comparebands}}
\end{figure*}

The B40 emission has a modest azimuthal asymmetry at 1.1 mm. The southwest side is brighter than the northeast side, with a $0.073\pm0.014$ mJy beam$^{-1}$ ($\sim8\%$) difference in peak intensities. \citet{2018ApJ...865...37M} report a similar brightness asymmetry at the 5$\sigma$ level in lower-resolution 930 $\mu$m data and at the $2\sigma$ level in 7 mm data. In our 2.1 mm data, the peak intensity of the northeast side is actually $\sim5\%$ brighter than the southwest side, but the difference is not statistically significant. The SNR of the 2.1 mm data is only about half that of the 1.1 mm data, making it more difficult to determine whether asymmetries are present.

\subsection{Continuum spectral index}\label{sec:spectralindex}
Measuring intensities at different frequencies can constrain the optical depth and dust grain properties of disks \citep[e.g.,][]{1991ApJ...381..250B}. To compare the GM Aur disk emission in different bands, the CASA \texttt{imsmooth} task is used to smooth the 1.1 mm continuum image to the same resolution as the 2.1 mm image (57 mas $\times$ 34 mas $(-13\fdg2)$). The normalized intensity profiles are shown in the top left of Figure \ref{fig:comparebands}. While the emission profiles are similar, there are several notable differences. At 2.1 mm, the emission rings are slightly narrower, the contrasts between the ring peaks and gap troughs are slightly larger, the central emission component is brighter relative to the peak intensity, and the relative brightness of the outer diffuse emission is lower.

Changes in intensity as a function of frequency are quantified with the spectral index, $\sfrac{d\log I_\nu}{d\log \nu}$. It is typically assumed that the intensity scales as $I_\nu \propto \nu^{\alpha}$, in which case

\begin{equation}
\alpha = \frac{\log{\sfrac{I_{\nu_1}}{I_{\nu_0}}}}{\log{\sfrac{\nu_1}{\nu_0}}}.
\end{equation}
Accounting for the $\sim10\%$ systematic flux calibration uncertainty in each band and assuming that the probability distribution is Gaussian, the disk-averaged spectral index between 1.1 and 2.1 mm is $\alpha=2.7\pm0.2$. This is similar to the disk-averaged spectral index ($2.94\pm0.44$) that \citet{2014AA...564A..51P} measured for GM Aur between 880 $\mu$m and 3 mm.  

There is significant radial variation in $\alpha$, as shown in the bottom left of Figure \ref{fig:comparebands}. While the systematic flux calibration uncertainty leads to a large uncertainty ($\sim0.2$, like that for the disk-averaged $\alpha$) in the absolute offset of the radial $\alpha$ profile, the uncertainty of the profile shape is much smaller because it is determined by the SNR of the observations. Interior to $R\sim100$ au, local minima in $\alpha$ occur near the emission peaks, while local maxima in $\alpha$ occur near continuum gaps. The $\alpha$ variations suggest that either larger dust grains are segregated in the rings or that the rings are optically thick while the gaps are optically thin \citep[e.g.,][]{2012AA...540A...6R, 2016ApJ...829L..35T}.  We examine the extent to which we can distinguish between these possibilities in Section \ref{sec:continuummodels}. 

In the outer diffuse emission region, the $\alpha$ profile measured from the high-resolution images is noisy. To improve sensitivity, we re-image both bands of data by applying a Gaussian taper of 0\farcs06 and then using \texttt{imsmooth} to smooth to a common resolution of 110 mas $\times$ 90 mas. The corresponding radial intensity and $\alpha$ profiles are shown on the right side of Figure \ref{fig:comparebands}. In the inner 100 au, the $\alpha$ variations are muted due to the degraded resolution, but in the outer disk, the improved sensitivity reveals modest radial variations in $\alpha$. In contrast with the pattern established in the inner disk, the local extrema do not coincide with the locations of D145 and B168.

\section{Continuum Models\label{sec:continuummodels}}
\subsection{Surface brightness models \label{sec:surfbrightness}}

\begin{deluxetable*}{ccccccc}
\tablecaption{Continuum surface brightness model parameters \label{tab:contmodel}}
\tablehead{
\colhead{Parameter}&\colhead{Prior type}&\colhead{Band 6 prior\tablenotemark{a}} &\colhead{Band 4 prior\tablenotemark{a}}& \colhead{Band 6 result}&\colhead{Band 4 result}&\colhead{Units}}
\startdata
$A_0$ & Uniform & [0, $2\times10^{-4}$]& [0, $2\times10^{-4}$] &$1.1\times10^{-4}\pm 9\times10^{-6}$& $8.9\times 10^{-5}\substack{+6\times 10^{-6}\\-7\times 10^{-6}}$ & Jy \\
$A_{1a}$ & Uniform &$[0.3,0.7]$& [0.1, 0.25] & $0.50 \pm 0.02$& $0.167\substack{+0.005\\-0.004}$ & Jy arcsec$^{-2}$\\
$r_{1a}$& Uniform &$[0.2, 0.25]$&[0.2, 0.25]& $0.2337 \pm 0.0006$&$0.2361\pm0.0003$& arcsec \\
$\sigma_{1a}$& Uniform &$[0.02, 0.06]$ & [0.01, 0.03] &$0.0257\substack{+ 0.0005\\-0.0006}$ & $0.0160\pm0.0009$& arcsec\\
$A_{1b}$ & Uniform &  [0.3, 0.7]& [0.05, 0.25]  &$0.456\pm0.007$& $0.131\substack{+0.002\\-0.003}$ & Jy arcsec$^{-2}$\\
$\Delta r$& Uniform &$[0, 0.1]$&[0, 0.1]&$0.0641\substack{+0.0011\\-0.0013}$&$0.049\pm0.002$& arcsec \\
$\sigma_{1b}$& Uniform &$[0.02, 0.06]$ & [0.02, 0.05] & $0.0444\substack{+0.0009\\-0.0008}$ & $0.0441\substack{+0.0009\\-0.0012}$& arcsec\\
$A_{2}$ & Uniform &[0.25, 0.3] & [0.05, 0.1] & $0.2778\pm0.0008$& $0.0812\pm0.0007$& Jy arcsec$^{-2}$\\
$r_{2}$& Uniform &$[0.5, 0.55]$& [0.5, 0.55] & $0.5202\pm0.0005$& $0.5222\pm0.0012$& arcsec \\
$\sigma_{2,in}$& Uniform &$[0.02, 0.06]$ & [0.02, 0.06]&$0.0278\pm0.0005$ & $0.025\pm0.001$& arcsec\\
$\sigma_{2,out}$& Uniform &$[0.02, 0.06]$ & [0.02, 0.06] &$0.0523\pm0.0004$ & $0.0449\pm0.0009$& arcsec\\
$A_{3}$ & Uniform &[0.025, 0.045] & [0, 0.01] & $0.0384\pm0.0002$ & $0.00597\substack{+0.0001\\-0.00009}$& Jy arcsec$^{-2}$\\
$r_{3}$& Uniform &$[0.4, 0.9]$& [0.4, 0.9] &$0.553\pm0.008$& $0.68\pm0.03$ & arcsec \\
$\sigma_{3}$& Uniform &$[0.35, 0.55]$ & [0.35, 0.55] &$0.495\pm0.003$ & $0.431\pm0.012$& arcsec\\
$A_{4}$ & Uniform &[0.005, 0.02]& [0, 0.005] &  $0.00954\pm0.00014$& $0.00182\pm0.00012$& Jy arcsec$^{-2}$\\
$r_{4}$& Uniform &$[1, 1.25]$&[1, 1.25]&$1.114\pm0.002$&$1.119\pm0.006$& arcsec \\
$\sigma_{4}$& Uniform &$[0.05, 0.2]$ & [0.05, 0.25] &$0.135\pm0.004$ & $0.097\pm0.012$ & arcsec\\
$i$  & Gaussian &[52.8, 1.5]& [52.8, 1.5]&$53.20\pm0.01$& $53.29\pm0.03$& degree\\
P. A. & Gaussian & [56.5, 2] & [56.5, 2]  & $57.16\pm0.02$&$57.24\pm0.05$ & degree\\
$\delta_x$ & Gaussian&[0, 0.02]&[0, 0.02] &$-0.00172\substack{+ 0.00005\\-0.00004}$ &$0.0072\pm0.0001$ & arcsec \\
$\delta_y$& Gaussian&[0, 0.02]&[0, 0.02] &$-0.00428\pm0.00005 $ &$-0.0011\pm0.0001$ & arcsec \\
\enddata
\tablenotetext{a}{If the prior is uniform, the numbers in brackets denote the bounds of the prior. If the prior is Gaussian, the first number corresponds to the center of the Gaussian and the second number corresponds to the standard deviation.}
\end{deluxetable*}

Interpreting the spectral index profile requires an estimate of the disk optical depth. To do this, we fit for GM Aur's surface brightness profile at each wavelength in the $uv$ plane. To make fitting the data more computationally tractable, each SPW is first frequency-averaged down to a single channel. Since CASA measurement sets store $uv$ coordinates in units of meters, we convert the coordinates to wavelength units ($\lambda$) using the frequencies of the individual SPWs. To reduce the data volume further, we follow the example of \citet{2016ApJ...823...37H} for modeling long-baseline ALMA data and bin the visibilities for each ALMA band into 12 k$\lambda\times$12 k$\lambda$ cells in the $uv$ plane (i.e., comparable at 1.1 mm to the ALMA antenna diameter).

An axisymmetric model is adopted, given that the azimuthal variations described in Section \ref{sec:results} are modest. The surface brightness profile is parametrized such that the central emission component is represented by a delta function, B40 is modeled as the sum of two overlapping Gaussian rings to account for the ``bump'' observed in Band 4, B84 is modeled as an ``asymmetric Gaussian'' ring, B168 is modeled as a Gaussian ring, and the diffuse outer emission is modeled as a broad Gaussian that is truncated at $r=0$. This profile can be written as 

\begin{align}\label{eq1}
I_\nu(r) &= A_0\int_{0}^{\infty} \delta(r') dr' \\
&+ A_{1a}\exp{\left(- \frac{(r-r_{1a})^2}{2\sigma_{1a}^2}\right)}\nonumber\\
&+ A_{1b}\exp{\left(- \frac{(r-r_{1a}-\Delta r)^2}{2\sigma_{1b}^2}\right)}\nonumber\\
&+ A_{2} \exp{\left(- \frac{(r-r_{2})^2}{2\sigma_2^2}\right)}\nonumber\\
&+\sum_{i=3}^{i=4} A_i\exp{\left(- \frac{(r-r_i)^2}{2\sigma_i^2}\right)}\nonumber
\end{align}

where 

\begin{align}
\sigma_2 = \left\{
        \begin{array}{ll}
          \sigma_{2,in}   & \quad r \leq r_2 \\
            \sigma_{2,out}& \quad r > r_2
        \end{array}
    \right.
\end{align}
This surface brightness prescription is partly motivated by \citet{2018ApJ...865...37M}, which used three concentric Gaussian rings to model lower resolution observations of GM Aur at 930 $\mu$m and 7 mm. The additional free parameters are the disk's P. A., $\cos i$, R. A. offset from the phase center ($\delta_\alpha$), and Decl. offset from the phase center ($\delta_\delta$), for a total of 21 free parameters. Positive offsets are defined to be north and east of the phase center, respectively. 

A model disk image is first generated without the central point source. Using the Python package \texttt{vis\_sample} \citep{2018AJ....155..182L}, synthetic visibilities $\mathcal{V}_m$ are produced by sampling the model image at the same $uv$ coordinates as the observations and performing a phase-shift. The central point source is directly added in the $uv$ plane in the form of a constant $A_0$ phase-shifted by the model offset. Each model is compared to the observed visibilities $\mathcal{V}_d$ using the log-likelihood function

\begin{equation}
\log p(\mathcal{V}_d|\Theta)=-\frac{1}{2}\sum_i \left( W_i |\mathcal{V}_{d,i}-\mathcal{V}_{m,i}|^2 + \ln \frac{2\pi}{W_i}\right),
\end{equation}
where $W_i$ is the weight corresponding to visibility $\mathcal{V}_i$ and $\Theta$ are the model parameters. The weights used in the likelihood calculations are scaled down from the nominal data weights provided in the delivered measurement sets by a factor of 2.667 because CASA's weight averaging procedure during data binning does not account for the effective channel width introduced by Hanning smoothing. This scaling factor was checked by computing the scatter of visibilities close to one another in $uv$ space.  

Uniform priors are adopted for the parameters defining $I_\nu$. The bounds of the priors were determined through a combination of considerations, including previous modeling results for GM Aur from \citet{2018ApJ...865...37M}, the size of the synthesized beam (i.e., lower or upper bounds for the widths of sources can be set depending on whether the features are well-resolved), and from manual testing to check that the priors are not overly restrictive. Gaussian priors are selected for the parameters governing the disk orientation and phase offset. The priors for the P. A. and $\cos (i)$ are set to the best-fit values derived in \citet{2018ApJ...865...37M}, while the standard deviations are set to be a few times wider than the posteriors from \citet{2018ApJ...865...37M} because additional substructures are being modeled and can affect the disk orientation measurements. The priors for $\delta_\alpha$ and $\delta_\delta$  are centered at 0, and each has a standard deviation of $0\farcs02$ (comparable to the scale of the synthesized beam). All priors are listed in Table \ref{tab:contmodel}.

The posterior probability distributions for the models at each wavelength are sampled with the \texttt{emcee} implementation of the affine invariant MCMC sampler \citep{2010CAMCS...5...65G, 2013PASP..125..306F}. Each ensemble employs 96 walkers for 20,000 steps, with the first 10,000 steps discarded as burn-in. Convergence is checked by verifying that the chains are much longer than the estimated autocorrelation times (typically on the order of a few hundred). The median values of the posterior distribution are listed in Table \ref{tab:contmodel}, with error bars computed from the 16th and 84th percentiles.

\begin{figure*}
\begin{center}
\includegraphics{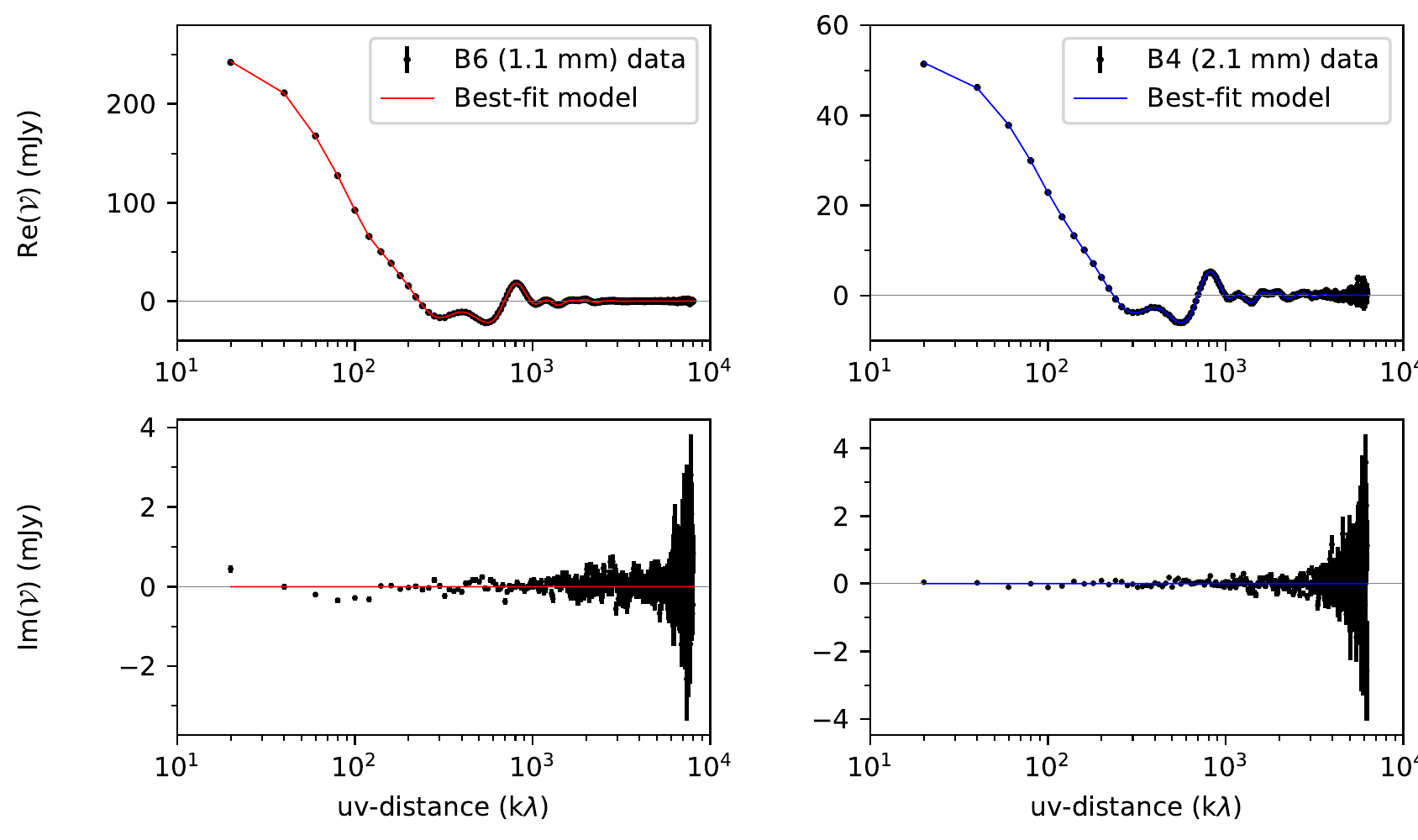}
\end{center}
\caption{\textit{Left column}: A comparison of the deprojected, azimuthally averaged Band 6 data and best-fit surface brightness model visibilities. The top row shows the real part of the visibilities and the bottom row shows the imaginary part. \textit{Right column:} Similar to the left column, but for Band 4. \label{fig:visibilitycomparison}}
\end{figure*}

The deprojected, azimuthally averaged visibilities corresponding to the best-fit models are compared to the observed visibilities in Figure \ref{fig:visibilitycomparison}. The models reproduce the real part of the visibilities well. Because the models are axisymmetric, the imaginary part of the deprojected and azimuthally averaged visibilities by construction should be zero (apart from numerical noise), but non-axisymmetric structure manifests in the data as non-zero imaginary components of the visibilities. The model and residual visibilities are then imaged with CASA in the same manner as the observations. A comparison of the model images and radial profiles to the data, as well as the residual images, are shown in Figure \ref{fig:modelimagecomparison}. Some significant residuals remain, as large as 8.4$\sigma$ at 1.1 mm and 6.8$\sigma$ at 2.1 mm. Part of the residuals are due to the non-axisymmetric structure around B40 discussed in Section \ref{sec:results}. The residuals around B84 have a systematic appearance, with the 1.1 mm model over-predicting emission along the major axis and under-predicting emission along the minor axis. Explanations for the discrepancy at B84 are discussed further in Section \ref{sec:RTmodels}. In addition, the B168 ring is not as pronounced in the 1.1 mm model as in the observations. However, the model radial profiles reproduce the observed intensities sufficiently well to for the purpose of examining optical depths in Section \ref{sec:dustmodels}. 

\begin{figure*}
\begin{center}
\includegraphics{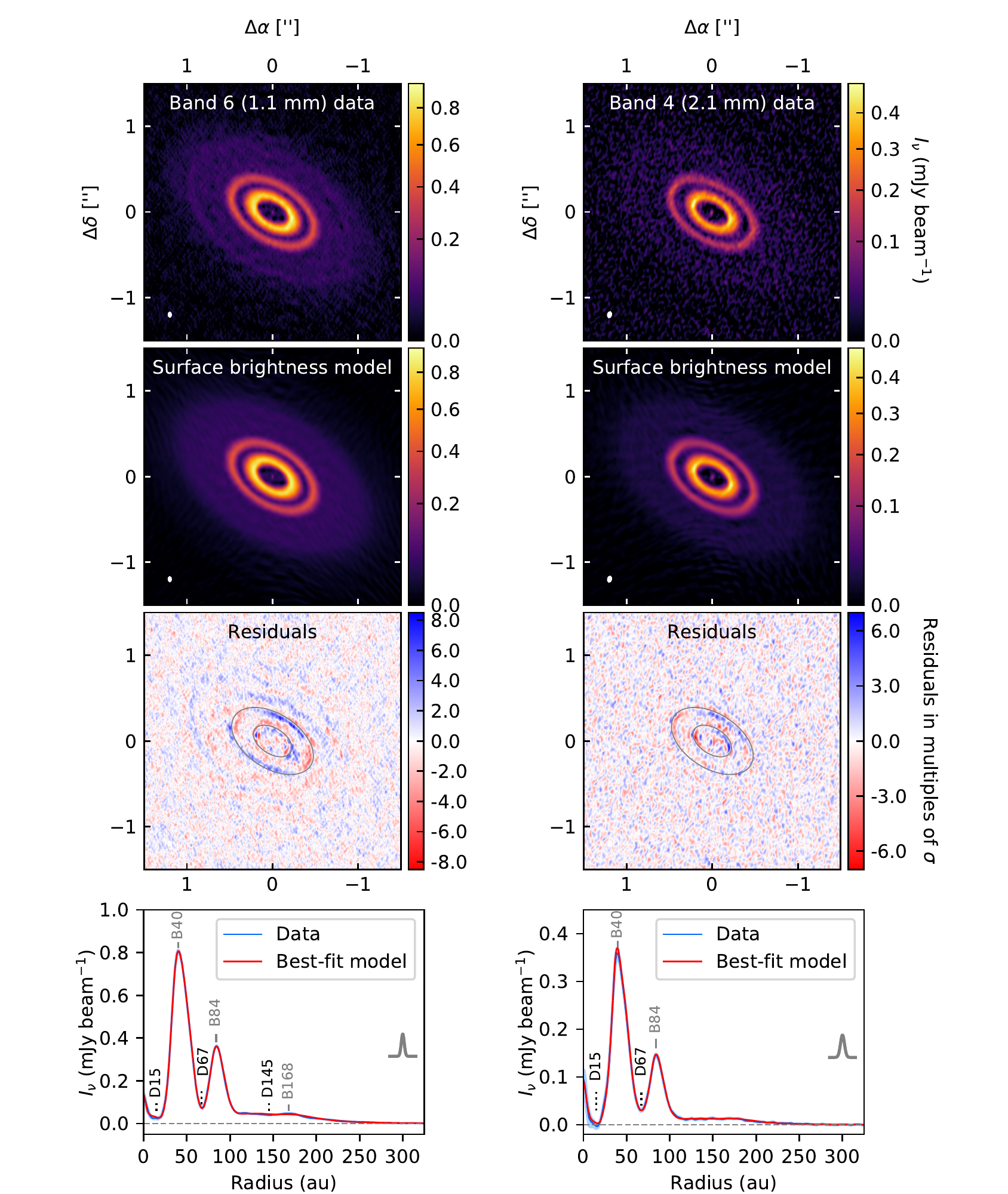}
\end{center}
\caption{\textit{Left column}: An image plane comparison of the Band 6 data and best-fit surface brightness model. The first row shows the CLEAN image of the data, the second row shows the CLEAN image of the model, and the third row shows the CLEAN image of the residuals scaled to the noise level $\sigma$. The gray ellipses mark the locations of B40 and B84. The fourth row compares the deprojected and azimuthally averaged radial profiles of the data and model, with the minor axis of the synthesized beam represented as a Gaussian on the right side of the plot. \textit{Right column:} Similar to the left column, but for Band 4. \label{fig:modelimagecomparison}}
\end{figure*}

\begin{figure}
\begin{center}
\includegraphics{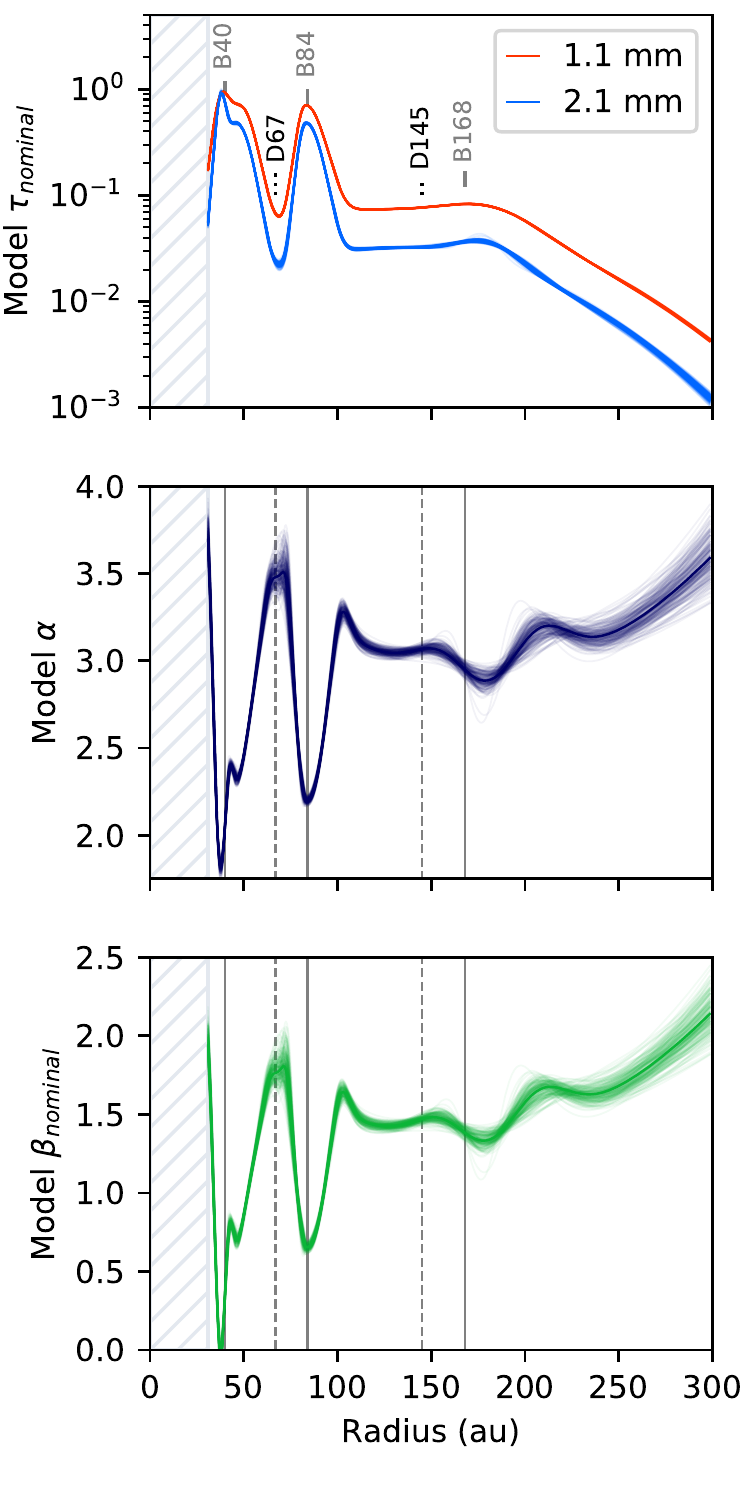}
\end{center}
\caption{\textit{Top:} Nominal optical depths corresponding to surface brightness profiles generated from 200 posterior samples, assuming scattering is negligible. Radii interior to 31 au are shaded because the temperature model from \citet{2018ApJ...865...37M} does not cover this region. \textit{Middle}: Spectral index profiles ($\alpha$) generated from the best-fit surface brightness profiles (dark purple) and 200 posterior samples (light purple). Solid lines mark the locations of emission rings and dashed lines denote gaps. \textit{Bottom:} Dust absorption opacity index ($\beta$) generated from the nominal optical depths shown in the top figure. The dark green curve is based on the best-fit surface brightness profile and the light green curves are based on the 200 posterior samples. \label{fig:modelopticaldepths}}
\end{figure}

\subsection{Constraints on dust properties \label{sec:dustmodels}}
\subsubsection{Constraints on optical depths}
We now use our model intensity profiles to determine whether optically thick emission can account for the low spectral index values of GM Aur's dust rings. To assess the optical depths, we first plot the quantity 
\begin{equation} \label{eq:taunominal}
\tau_\text{nominal}(r)=-\ln\left(1-\frac{I_\nu(r)}{B_\nu(r)}\right)
\end{equation}
in Figure \ref{fig:modelopticaldepths}. $I_\nu$ is the surface brightness model at a given frequency $\nu$ and B$_\nu$ is the Planck function evaluated at the midplane dust temperatures derived in \citet{2018ApJ...865...37M} through radiative transfer modeling of GM Aur's SED and resolved millimeter continuum observations. The expression for $\tau_\text{nominal}$ is typically used to estimate the optical depth in the limit where the dust is optically thin or the scattering opacities are small. The spectral index profile $\alpha$ computed from the best-fit surface brightness models is plotted in the same figure. 

The dominant source of uncertainty in $\tau_\text{nominal}$ is the midplane dust temperature. Unfortunately, the uncertainties associated with dust temperatures derived from radiative transfer modeling are usually ill-quantified due to the computational expense of exploring parameter space. However, the 1.1 mm continuum brightness temperatures set a lower bound on the possible midplane dust temperatures\textemdash the true midplane temperatures cannot be lower than the model temperatures by more than $\sim35\%$. As shown later in Section \ref{sec:hcop}, the brightness temperatures of optically thick HCO$^+$, which emits from a warmer elevated layer, indicate that the true midplane temperatures cannot be more than a factor of two higher than the model temperatures.

With these uncertainties in mind, we can use Figure \ref{fig:modelopticaldepths} to examine which parts of the disk are likely to be optically thick or thin. $\tau_\text{nominal}<1$ throughout the disk at both wavelengths. However, one cannot immediately conclude that the disk is completely optically thin\textemdash the peak values at B40 and B84 at 1.1 mm are high enough ($\tau_\text{nominal}=0.94$ and $0.70$, respectively) such that assuming midplane temperatures a few degrees lower would push the values of $\tau_\text{nominal}$ above 1. Furthermore, \citet{2019ApJ...877L..18Z} point out that when the dust albedo is sufficiently high, $\tau_\text{nominal}$  can be as low as $\sim0.6$ at millimeter wavelengths even in an optically thick disk. 

On the other hand, $\tau_\text{nominal}\ll1$ at D67 and beyond $R\sim100$ au. Even with large temperature uncertainties, these regions must be optically thin and therefore $\tau_\text{nominal}$ is a good approximation of the optical depth. The low optical depths are expected given the high spectral indices ($\alpha\gtrapprox3$) in these regions. The gap at D67 is not completely evacuated\textemdash the best fit model indicates $\tau_\text{1.1 mm}\sim0.06$. Although the inner boundary of the temperature model stops short of the innermost gap, the intensities interior to 30 au are lower than the intensity at D67, which is presumably colder than the inner disk. Thus, the innermost gap should also be optically thin (with perhaps the exception of the central unresolved emission, which is discussed in Section \ref{sec:innerdisk}). 

\begin{figure}
\begin{center}
\includegraphics{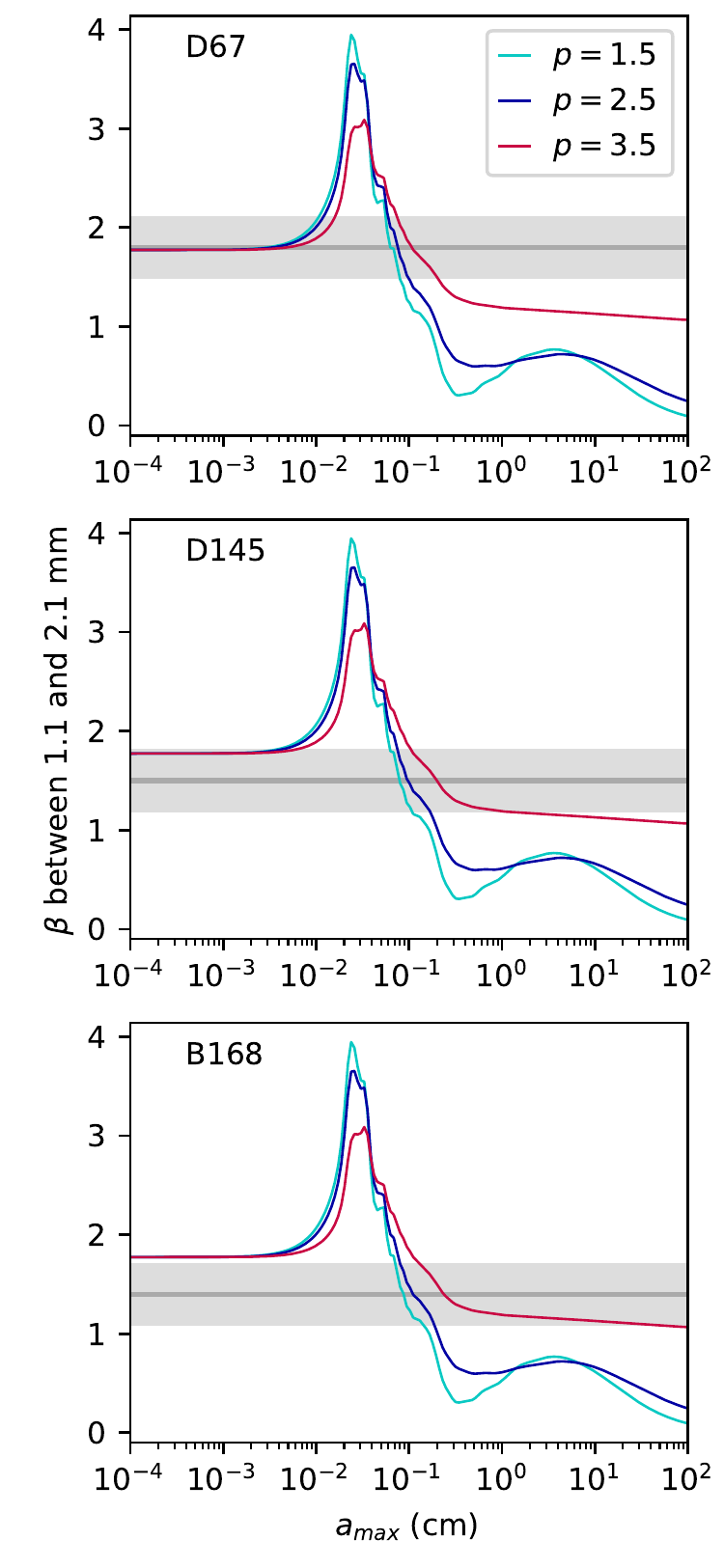}
\end{center}
\caption{A comparison of the dust absorption opacity index $\beta$ between 1.1 and 2.1 mm to the $\beta_\text{nominal}$ values derived for GM Aur. The $\beta$ curves are computed as a function of $a_\text{max}$ and $p$ using the ``DSHARP dust opacities'' from \citet{2018ApJ...869L..45B}. From top to bottom, the dark gray horizontal lines show the $\beta$ values from the best-fit GM Aur model at D67, D145, and B168, respectively. The light gray region shows the possible range of values after taking the systematic flux calibration uncertainty into account. The intersection between the gray regions and the colored curves indicate which $a_\text{max}$ values are consistent with the $\beta$ measurements for GM Aur.  \label{fig:DSHARPbeta}}
\end{figure}

\begin{figure*}
\begin{center}
\includegraphics{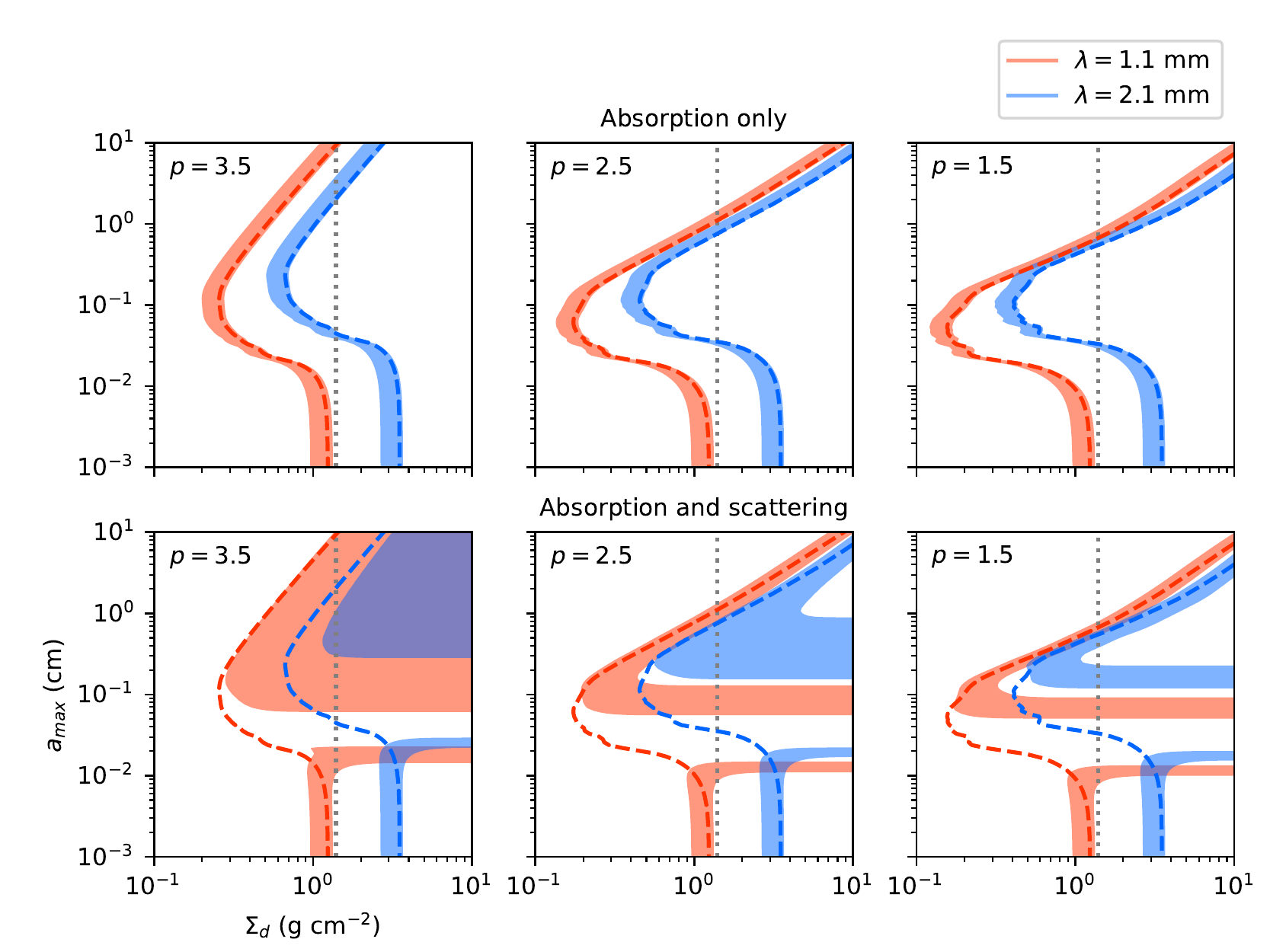}
\end{center}
\caption{Plots showing which combinations of $\Sigma_d$ and $a_\text{max}$ yield intensity values within 10\% of the best-fit model intensities at each wavelength at the peak of B40. Overlapping shaded regions indicate which parameter combinations are consistent with both sets of continuum observations. The grids are computed for $T_d=30$ K. As an optical depth reference, the dashed lines mark where $\tau_\text{abs}=\kappa_\text{abs}\Sigma_d/\mu=1$. The gray dotted line marks the value of $\Sigma_d$ at which the Toomre $Q$ parameter is 1, assuming a gas-to-dust ratio of 100. \textit{Top}: Results when scattering opacities are set to zero everywhere. \textit{Bottom}: Results accounting for both absorption and scattering. Accounting for scattering shows that relatively small grain sizes can reproduce the observed intensities, but does not eliminate the possibility of large grains being present. \label{fig:B40scattering}}
\end{figure*}

\begin{figure*}
\begin{center}
\includegraphics{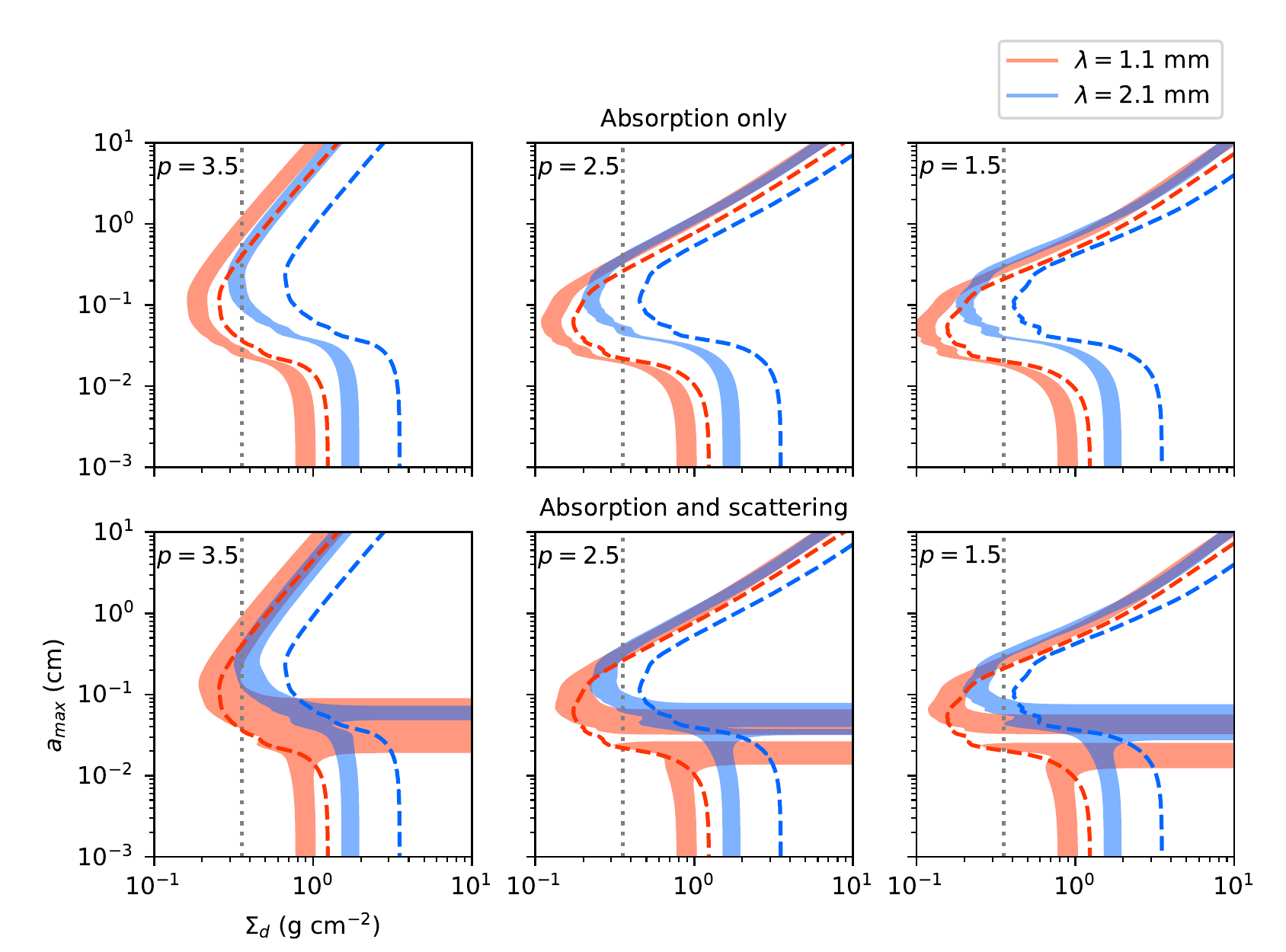}
\end{center}
\caption{Similar to Figure \ref{fig:B40scattering}, but for the intensity values at the B84 emission ring. The grids are computed for $T_d = 18$ K. \label{fig:B84scattering}}
\end{figure*}

\subsubsection{Constraints on dust grain sizes}
For optically thin disk regions, dust grain sizes can be constrained by measuring how the optical depth changes between two frequencies. Because of the uncertainties associated with disk temperature and absolute flux calibration, the aim of this section is to comment on which regions of parameter space are consistent with the observations rather than to provide precise measurements. 

The frequency dependence of the dust absorption opacity $\kappa_\text{abs}$ is usually quantified with the $\beta$ parameter, where $\kappa_\text{abs} \propto \nu^\beta$. Since $\tau_\text{abs}=\sfrac{\kappa_\text{abs}\Sigma_d}{\mu}$, where $\Sigma_d$ is the dust surface density and $\mu=\cos i$, $\tau_\text{abs}$ has the same frequency dependence as $\kappa_\text{abs}$. Thus we can use the $\tau_\text{nominal}$ values measured from the 1.1 and 2.1 mm surface brightness models to estimate $\beta$ in optically thin parts of the disk. The model $\beta$ radial profile (labeled $\beta_\text{nominal}$ to signify that it is computed by neglecting scattering) is shown in Figure \ref{fig:modelopticaldepths}. 

To connect $\beta$ values to grain sizes, we adopt the default ``DSHARP dust opacities'' described in \citet{2018ApJ...869L..45B} and use the companion Python package \texttt{dsharp\_opac} \citep{2018zndo...1495277B} to compute quantities derived from the opacities. The opacities are based on optical constants from \citet{1996AA...311..291H}, \citet{2003ARAA..41..241D}, and \citet{2008JGRD..11314220W}. Throughout this paper, we assume the dust population follows a power-law size distribution $n(a)\propto a^{-p}$ and fix the minimum grain size at 0.1 $\mu$m. The specific choice of $a_\text{min}$ does not have a large effect on the millimeter wavelength opacities as long as it is much smaller than $a_\text{max}$ \citep[e.g.,][]{2006ApJ...636.1114D}. Figure \ref{fig:DSHARPbeta} shows how $\beta$ varies as a function of the maximum dust grain size $a_\text{max}$ for three different power laws: $p=3.5$ (i.e., the standard ISM value from \citet{1977ApJ...217..425M}), $p=2.5$, and $p=1.5$. The shallower power-law distributions may arise via grain growth \citep[e.g.,][]{1997Icar..128..429W}. The $\beta$ values measured for GM Aur at D67, D145, and B168 are shaded in gray. They are best matched by $a_\text{max}$ values from $1-3$ mm, with D67 also matching well with $a_\text{max}$ values smaller than $\sim$ 100 $\mu$m due to the non-monotonic shape of $\beta$ as a function of $a_\text{max}$. However, the 10\% absolute flux calibration uncertainty leads to an absolute offset uncertainty of $\sim0.3$ in GM Aur's $\beta$ profile, so the $\beta$ values at these locations are consistent with a large range of grain sizes. Nevertheless, the upward trend in $\beta$ beyond 200 au indicates that it is unlikely that  $a_\text{max}$ is significantly less than 1 mm in the optically thin outer disk ($R\gtrsim100$ au).  The very high $\beta$ values near $a_\text{max}\sim0.5$ mm are inconsistent with the observations. For values of $a_\text{max}\lesssim 0.5$ mm, $\beta$ is either flat or an increasing function of $a_\text{max}$. In this regime, GM Aur's $\beta$ profile in the outer disk would imply that $a_\text{max}$ is increasing with distance from the star, which would be difficult to reconcile with standard models indicating that larger dust grains preferentially drift inward \citep[e.g.,][]{1977ApSS..51..153W}.  

The estimated peak optical depths of B40 and B84 are high enough that using $\beta_\text{nominal}$ to constrain grain sizes can lead to significant over-estimates \citep[e.g.,][]{2019ApJ...883...71C}. Instead, the effects of scattering should explicitly be considered. To compute the emergent intensity $I_\nu^\text{out}$, we use the analytic approximations from  \citet{2019ApJ...876....7S} and \citet{2019ApJ...883...71C}, which are summarized in Appendix \ref{sec:scatteringformula}. Similar results are obtained using formulae based on the Eddington-Barbier approximation from \citet{2018ApJ...869L..45B} and \citet{2019ApJ...877L..18Z}. As noted in the appendix, $I_\nu^\text{out}$ depends on five parameters: $p$, $a_\text{max}$, the dust temperature $T_d$, $\Sigma_d$, and $\mu=\cos i$. The disk inclination is known, and we can use the midplane dust temperatures derived in \citet{2018ApJ...865...37M}. This still leaves three free parameters with only two constraints (i.e., the intensities at each wavelength), so we cannot solve for the values of these parameters. Although GM Aur has been observed at other wavelengths \citep[e.g.,][]{2018ApJ...865...37M}, the much coarser spatial resolution of earlier observations prevents us from accurately estimating the intensities at the ring peaks. 

We can, however, examine how accounting for scattering affects inferences about $a_\text{max}$ and $\Sigma_d$ for several possible values of $p$. For each of $p=3.5$, $2.5$, and $1.5$, we compute the emergent intensities at 2.1 mm and 1.1 mm for a grid of $\Sigma_d$ and $a_{max}$ values at the temperatures corresponding to the peaks of B40 and B84.  One set of calculations (``Absorption only") is performed by setting the scattering opacities to zero everywhere, while ``Absorption and scattering" uses the DSHARP scattering opacities. To account for the flux calibration uncertainty, Figure \ref{fig:B40scattering} shades in the combinations of $a_\text{max}$ and $\Sigma_d$ that produce intensities within 10\% of the best fit model intensities at each wavelength at the peak of B40. Figure \ref{fig:B84scattering} does the same for B84. Overlapping shaded regions indicate which combinations of $a_\text{max}$ and $\Sigma_d$ are consistent with both bands of the GM Aur observations (leaving aside temperature uncertainties, since our focus is on the qualitative effect of neglecting scattering opacities). 

Comparing the two rows in Figure \ref{fig:B40scattering} shows that accounting for scattering allows for steeper $p$ values than would be inferred from absorption-only calculations. Both Figures \ref{fig:B40scattering} and \ref{fig:B84scattering} also demonstrate that accounting for scattering yields solutions for $a_\text{max}$ that can be smaller and for $\Sigma_d$ that can be larger than the solutions derived from absorption-only calculations. However, it is also important to note that the solution space can be discontinuous, and factoring in scattering does not necessarily rule out optically thin dust or grains larger than a millimeter at the ring peaks. For some values of $p$, the solution space becomes discontinuous somewhere between $a_\text{max}\sim0.01$ to $0.1$ cm (i.e., at values comparable to the wavelength of the observations) because the albedos are very high for these size distributions, so optically thick emission saturates below the measured intensities. This effect is explained in detail in \citet{2019ApJ...877L..18Z}. Thus, given the available data, it is ambiguous the extent to which trapping of large dust grains contributes to the low spectral index values measured at B40 and B84. 
 
Despite the aforementioned ambiguities, we argue that GM Aur's millimeter continuum emission is likely tracing some degree of radial variation in dust properties inside $R\sim100$ au. If at least one of the rings is optically thin, then the large spectral index variations across the ring(s) must be a consequence of dust opacity variations due to changing grain sizes or compositions \citep[e.g.,][]{1991ApJ...381..250B, 2003AA...403..323T}. If both rings are optically thick and the dust properties are radially uniform, then B84 would have a slightly lower spectral index than B40 due to the lower temperature at a larger radius (note that spectral indices are temperature dependent outside the Rayleigh-Jeans regime). Instead, the spectral index of B84 is higher. Thus, even if both rings are optically thick, the spectral index indicates that grains in B40 have different properties from the grains in B84. On the other hand, since the outer disk ($R\gtrsim100$ au) is optically thin, the spectral index variations point to radial variations in dust properties. However, the local minima and maxima in the spectral index profile do not coincide neatly with D145 and B168, so it is not clear whether B168 is a dust trap. 

Despite the large range of $a_\text{max}$ values consistent with GM Aur's ring intensities, Figures \ref{fig:B40scattering} and \ref{fig:B84scattering}  also show that for the largest and smallest $a_\text{max}$ solutions, the corresponding $\Sigma_d$ solutions are well above the surface density limit for which the Toomre $Q$ parameter is equal to 1, assuming the standard gas-to-dust ratio of 100 and a stellar mass of 1.32 M$_\odot$ \citep{2018ApJ...865..157A}. The continuum emission does not exhibit any clear signatures of gravitational instability, such as spiral arms, suggesting that $Q$ is greater than 1 at the emission rings. Thus, the gas-to-dust ratios at the rings would have to be below 100 if  $a_\text{max}$ is smaller than $\sim1$ mm or larger than a few millimeters. 

Of course, as illustrated in \citet{2018ApJ...869L..45B}, different assumptions about the dust grain composition and structure will affect how multi-wavelength intensities are translated to grain sizes. For example, unlike the compact DSHARP grains, the $\beta$ curves of highly porous grains are nearly flat as a function of $a_\text{max}$ \citep[e.g.,][]{2014AA...568A..42K}. In this case, the strong spectral index variations in the GM Aur disk indicate that the dust cannot simultaneously be optically thin and highly porous.

\begin{figure*}
\begin{center}
\includegraphics{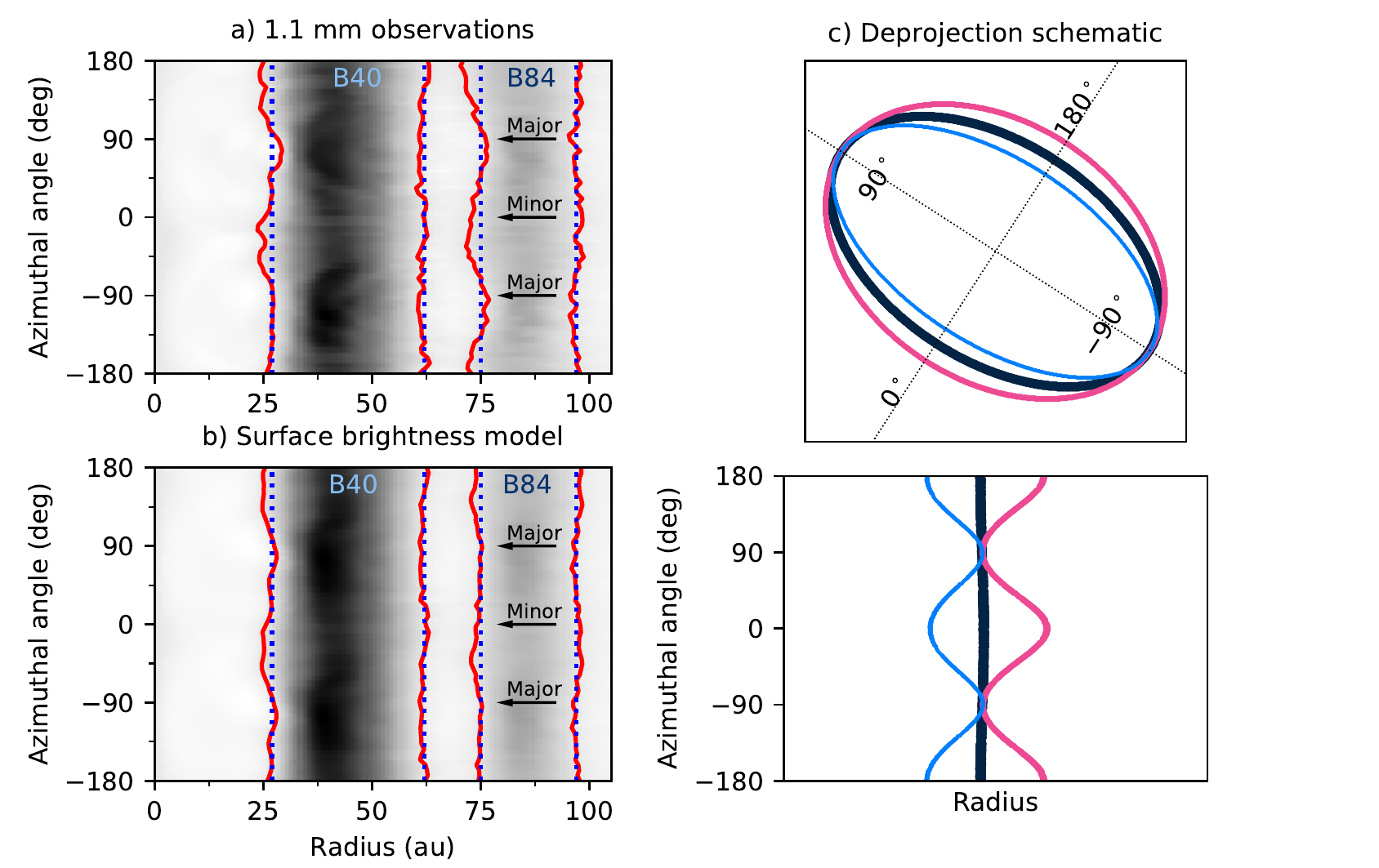}
\end{center}
\caption{\textit{a)} The 1.1 mm GM Aur continuum deprojected and replotted as a function of radius and azimuthal angle. The arrows show the azimuthal angles corresponding to the major and minor axes of GM Aur. The 15$\sigma$ contours are shown in red. Blue dotted vertical lines are drawn for reference to demonstrate that the inner contour of B84 in the observations appears to have a larger radius near azimuthal angles corresponding to the projected disk major axis ($\pm90^\circ$) than near angles corresponding to the  minor axis (0$^\circ$, $\pm180^\circ$). Meanwhile, the contours of B40 and the outer contour of B84 appear to be at constant radius. \textit{b)} Similar to \textit{a)}, but for the best-fit surface brightness model. All contours appear to be at constant radius, as expected for an axisymmetric model. \textit{c)} A schematic of how deprojected polar plots of rings with different inclinations would appear for a particular choice of deprojection geometry. The black ring demonstrates the case where the deprojection inclination matches that of the ring, so the black ring appears at constant radius on the polar plot. The blue ring demonstrates that if the inclination angle used for deprojection is smaller than that of the ring, the ring will trace a curve on the polar plot that has larger radial values at angles corresponding to the projected major axis compared to the minor axis. The pink ring demonstrates that if the inclination angle used for deprojection is larger than that of the ring, the ring will trace a curve on the polar plot that has larger radial values at angles corresponding to the projected minor axis compared to the major axis. \label{fig:B84comparison}}
\end{figure*}

\subsection{The geometry of B84}\label{sec:RTmodels}
Our surface brightness model setup is appropriate for axisymmetric emission originating from a geometrically thin layer. The residuals remaining in Figure \ref{fig:modelimagecomparison} suggest that at least one of these assumptions is not wholly appropriate. Examining the breakdown of these assumptions can provide useful clues into disk structure and evolution. The vertical distribution of dust traces how readily large grains settle to the midplane \citep[e.g.,][]{2016ApJ...816...25P}, while departures from axisymmetry may point to the presence of a perturbing body \citep[e.g.,][]{2001ApJ...560..997L}. 

The 1.1 mm residuals near B84 are particularly interesting due to their systematic appearance. As noted in Section \ref{sec:surfbrightness}, the surface brightness model overpredicts emission along the projected disk major axis and underpredicts along the minor axis, suggesting that the inner edge of B84 is slightly more elongated along the major axis compared to the outer edge. This is more readily seen by deprojecting the observations, replotting the continuum as a function of radius and azimuthal angle, and drawing the 15$\sigma$ contours to highlight the edges of B40 and B84 (Figure \ref{fig:B84comparison}). The major axis of the original image runs between 90 and $-90^\circ$, while the minor axis runs between $0$ and $180/-180^\circ$. The contours of B40 and the outer contour of B84 appear to be approximately vertical, indicating that they are at constant radius (i.e., they are axisymmetric). However, the inner contour of B84 undulates, tracing larger radii at azimuthal angles corresponding to the major axis than at angles corresponding to the minor axis. In other words, the inner edge of B84 appears to be at a higher inclination than the outer edge of the ring. (Choosing slightly higher or lower contour levels yields the same effect). For comparison, Figure \ref{fig:B84comparison} also shows a polar plot for the best-fit surface brightness model. All the contours appear to be at a constant radius, showing that the distorted inner edge of B84 is not a consequence of the $uv$ sampling or imaging artifacts.

To explore possible explanations for the geometry of B84, we use RADMC-3D \citep{2012ascl.soft02015D} to perform 3D Monte Carlo radiative transfer calculations and generate synthetic observations of three parametric ring models. As a reference, we first generate an optically and geometrically thin disk model (``Flat RT Model''). Since annular substructures often appear to have different inclinations in scattered light observations due to the projection of the optically thick, flared disk surface \citep[e.g.,][]{2016AA...595A.114D}, we next generate an optically and geometrically thick model (``Thick RT Model'') to determine whether a similar effect could be at play for GM Aur's millimeter continuum emission. Finally, we generate a mildly warped model (``Warped RT model'') based on previous hypotheses that the GM Aur disk is warped \citep[e.g.,][]{2008AA...490L..15D, 2009ApJ...698..131H}. 

The dust surface density radial profile is modeled as an asymmetric Gaussian ring, analogous to the surface brightness model profile for the B84 ring:

\begin{align}
\Sigma_d(r)  = \left\{
        \begin{array}{ll}
          C\exp{\left(-\frac{(r-84\,\text{au})^2}{2w_\text{in}^2}\right)}   & \quad r \leq 84\,\text{au} \\
            C\exp{\left(- \frac{(r-84\,\text{au})^2}{2w_\text{out}^2}\right)} & \quad r > 84\,\text{au}
        \end{array}
    \right.
\end{align}

The temperature is parametrized as 
\begin{equation}
T(r) = T_{10}\left(\frac{r}{\text{10 au}}\right)^{-q}
\end{equation}
We adopt a vertically isothermal temperature structure because the millimeter continuum emission presumably originates from large dust grains settled in the midplane, and therefore the temperature variation should not be large within this layer. The vertical distribution of the dust is Gaussian, so the dust density is given by 
\begin{equation}
\rho_d(r, z) = \frac{\Sigma_d(r)}{\sqrt{2\pi} h(r)}\exp\left(-\frac{z^2}{2h(r)^2}\right),
\end{equation}
where $h(r)$ is the dust scale height. Due to vertical settling, the dust scale height is assumed to be some constant fraction of the gas pressure scale height $H(r) = \sqrt{\frac{k_B T(r) r^3}{\mu_\text{gas}m_HGM_\ast}}$, where the mean molecular weight is $\mu m_H=2.37\times$the mass of atomic hydrogen. We use a stellar mass of 1.32 M$_\odot$ computed from stellar evolutionary tracks \citep{2018ApJ...865..157A}. 

We use a dust population with $p=3.5$ and $a_\text{max}=1$ mm, which corresponds to an absorption opacity of $\kappa_\text{abs}=2.4$ cm$^2$ g$^{-1}$ and scattering opacity of $\kappa_\text{sca}=20.6$ cm$^2$ g$^{-1}$ at 1.1 mm. A more realistic ``two-population'' dust model would also include a population of dust grains with a sub-micron $a_\text{max}$ in the upper layers of the disk \citep[e.g.,][]{2006ApJ...638..314D}, but the grains in the upper layer are not expected to contribute significantly to the millimeter continuum emission because they only constitute a small fraction of the solid mass. The small grains in the upper layers are important for self-consistent thermal structure calculations, but we parametrize our temperature structures because ALMA observations only constrain the properties of dust in the midplane. We note that Section \ref{sec:dustmodels} suggests that the dust properties in the GM Aur disk could be quite different from what we use in this section. Dust opacities, temperatures, and surface densities are highly degenerate when modeling continuum emission. Thus the models we present should be taken as illustrative, not as a ``best fit'' to GM Aur.

For each model, we compute the emission at 1.1 mm using 512 radial cells spaced logarithmically from 0.5 to 175 au, 1024 poloidal cells spaced evenly between $\frac{\pi}{6}$ and $\frac{5\pi}{6}$ (the midplane is at $\frac{\pi}{2}$), 64 azimuthal cells spaced evenly between 0 and $2\pi$, and $10^8$ photon packages. Anisotropic scattering is treated using M\"uller matrices calculated with the Draine version\footnote{\url{http://scatterlib.wikidot.com/mie}} of the Mie code by \citet{1983asls.book.....B}. The phase offset is fixed to the best-fit values derived from the 1.1 mm surface brightness modeling, while the overall disk inclination and P. A. are fixed to the weighted averages of the best-fit 1.1 and 2.1 mm models ($53\fdg21$ and $57\fdg17$, respectively). Synthetic visibilities are generated from the radiative transfer output using \texttt{vis\_sample}, and the resulting visibilities are imaged with CASA. The model parameters are listed in Table \ref{tab:RTmodel}. 

\begin{deluxetable}{llll}
\tablecaption{Continuum radiative transfer model parameters\label{tab:RTmodel}}
\tablehead{
\colhead{Parameter}&\colhead{Flat} &\colhead{Thick}& \colhead{Warped}}
\startdata
$T_{10}$ (K) &58.1&45&58.1\\
$q$& 0.525&0.525&0.525\\
$h(r)$ (au) &0.2$H(r)$&0.4$H(r)$&0.2$H(r)$\\
$C$ (g cm$^{-2}$) & 0.175 & 1.5&0.175\\
$w_{in}$ (au) & 5&3&3.5\\
$w_{out}$ (au) & 7.5 &4.5&7.5\\
$\Delta i_\text{max}$ (\textdegree) & -&- &5\\
\enddata
\end{deluxetable}

\begin{figure*}
\begin{center}
\includegraphics{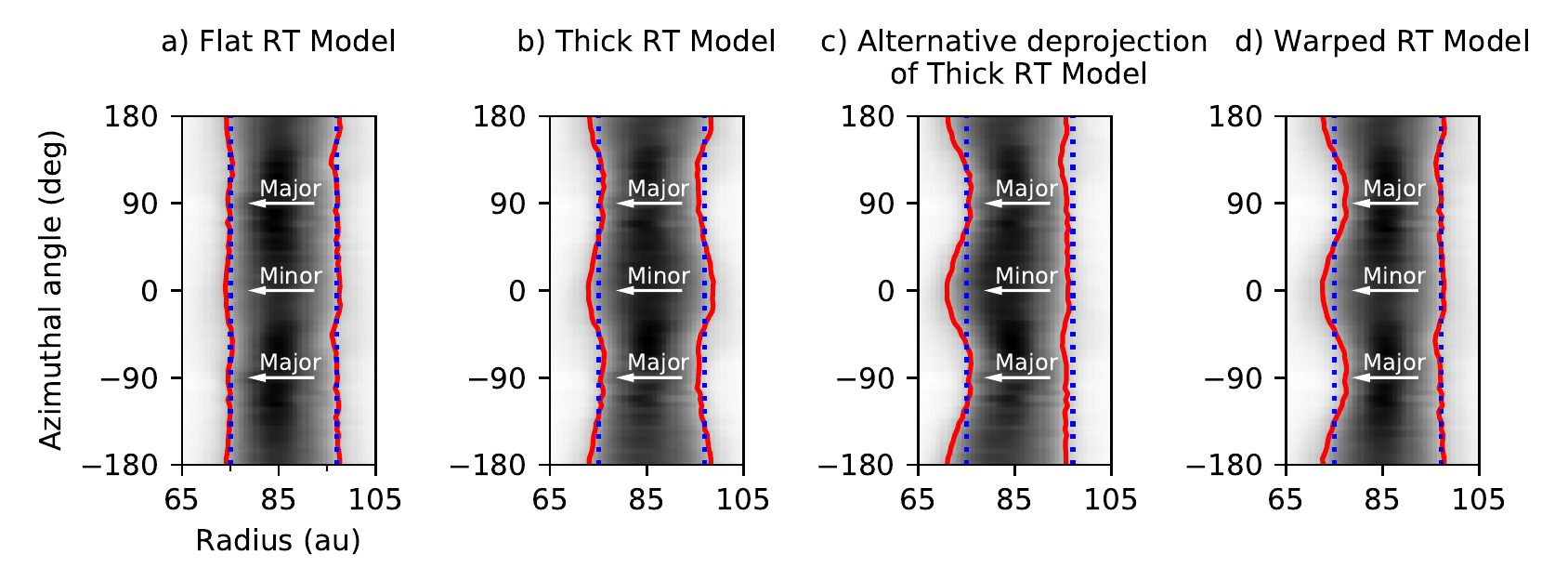}
\end{center}
\caption{Deprojected polar plots of radiative transfer models exploring different scenarios that might yield the appearance of changing orientation across the B84 ring. For all plots except c), a P. A. P. A.of $57\fdg17$ and $i=53\fdg21$ are used to deproject. The arrows show the azimuthal angles corresponding to the major and minor axes of GM Aur. Red contours are drawn at an intensity level of 0.15 mJy beam $^{-1}$ (equal to 15$\sigma$ in the 1.1 mm observations). Blue dotted vertical lines are drawn as a reference for constant radius. \textit{a)} Plot of a radiative transfer model of a geometrically and optically thin ring. The ring contours are approximately vertical (i.e., at constant radius). \textit{b)} Plot of a radiative transfer model of a geometrically and optically thick ring. The inner contour traces larger radii near azimuthal angles corresponding to the major axis of the disk image ($\pm90^\circ$) than near angles corresponding to the  minor axis (0$^\circ$, $\pm180^\circ$). Meanwhile, the outer contour has the opposite behavior. \textit{c)} Plot of the same model from \textit{b)}, except an inclination angle of 52$^\circ$ is used to deproject rather than the true inclination of $53\fdg21$. With this alternative deprojection, the outer contour now appears to be at constant radius, while the inner contour still undulates. \textit{d)} Plot of a radiative transfer model of a warped ring. The inner contour undulates similarly to the inner contour of B84 in the observations, while the outer contour is approximately vertical. \label{fig:RTmodels}}
\end{figure*}

The ``Flat RT Model'' is relatively settled: $h(r)=0.2H(r)$. The temperature structure is set by a power-law fit to the GM Aur midplane temperature calculated by \citet{2018ApJ...865...37M}. At the peak of the ring, the dust aspect ratio is $h/r~\sim0.014$. The parameters to generate the disk surface density are adjusted manually until the width and height of the ring in the CASA image of the radiative transfer model is comparable to that of B84. As shown in Figure \ref{fig:RTmodels}a, the contours trace approximately constant radii in the polar plot, as expected for a geometrically and optically thin disk. Doubling the dust scale height of the ``Flat RT'' (optically thin) model does not appreciably change the emission geometry of B84.

The emitting surface of the ``Thick RT Model'' is elevated by changing the dust scale height to $h(r)=0.4H(r)$ and increasing the surface density relative to the ``Flat RT Model'' such that B84 becomes optically thick. Because increasing the surface density also increases the intensity, we compensate by decreasing $T_\text{10}$ and the width of the ring until the width and height of the ring in the CASA image of the radiative transfer model is comparable to that of B84 in GM Aur. At the peak of the ring, the dust disk aspect ratio is $h/r\sim0.024$. Figure \ref{fig:RTmodels}b shows that similarly to the observations of B84, the inner contour of the radiative transfer model traces larger radii near azimuthal angles corresponding to the major axis of the disk image than near angles corresponding to the  minor axis. Thus, the apparent inclination of the inner edge is higher than the true inclination. Meanwhile, the outer contour also undulates, but in the opposite direction, indicating that the apparent inclination of the outer edge is lower than the true inclination. Figure \ref{fig:RTmodels}c demonstrates that by deprojecting the same model image with an inclination of $52^\circ$ rather than the true inclination of $53\fdg21$, one can obtain a polar plot where the outer contour appears straight on the polar plot but the inner contour undulates, similar to the GM Aur observations.

To produce the ``Warped RT model'', we start with the parameters from the ``Flat RT model'' and rotate dust annuli out of the disk plane by some angle $\Delta i$ around the projected disk major axis. The rotation angle is parametrized as 

\begin{align}
\Delta i(r) = \left\{
        \begin{array}{ll}
           \Delta i_\text{max}\left( 1-\frac{(r-75\text{ au})}{10 \text{ au}}\right)  &  \quad  r\in (75 \text{ au},\, 85 \text{ au}) \\
            0& \quad \text{everywhere else}
        \end{array}
    \right.
\end{align}
In this parametrization, the inner edge of B84 is mildly misaligned with the plane of the disk, while the outer edge is coplanar. The radiative transfer models are generated such that the inner edge appears more highly inclined than the overall disk from the viewer's vantage point. The parameters are again adjusted until the width of the ring in the CASA image of the radiative transfer model is comparable to that of B84 in GM Aur. As shown in Figure \ref{fig:RTmodels}, a modest value of $i_\text{max}=5$\textdegree\,can yield an undulating inner contour and a straight outer contour, mimicking the behavior of the B84 ring.

The radiative transfer models indicate that non-axisymmetric structure and vertical structure are both plausible explanations for the emission geometry of B84. Better sensitivity and resolution could help to distinguish between these and other scenarios. Along the projected minor axis of the disk, the optically and geometrically thick radiative transfer model produces emission that is $\sim10\%$ brighter on the far side (southeast) of the B84 ring compared to the near (northwest) side because the warmer inner rim of the far side is more exposed to the viewer. The SNR of the B84 emission in the current GM Aur observations is not high enough to confirm whether such a difference is present (note that the $\sim8\%$ brightness difference for B40 along the projected major axis, as described in Section \ref{sec:results}, is statistically significant because the inner ring is much brighter). Meanwhile, although our warped model is set up such that the inner edge of B84 is tilted around the projected disk major axis for the sake of simplicity, it is unlikely that GM Aur would have such a coincidental orientation. The case for a warp could be strengthened if better quality observations also demonstrated that the P. A. of the inner edge of B84 differs from that of the outer edge. High spectral and spatial resolution line observations can also be used to test for the presence of a warp \citep[e.g.][]{2014ApJ...782...62R}, although warps do not always leave a clear imprint on observed disk kinematics \citep[e.g.,][]{2017MNRAS.466.4053J}. Thus, hydrodynamical simulations with more realistic warp geometries will also be needed to determine whether the observations are compatible with a disk warp. It should also be noted that if the radial thermal profile is more complex than assumed due to additional heating or cooling in the gaps \citep[e.g.,][]{2018AA...612A.104F, 2018ApJ...867L..14V}, fully self-consistent radiative transfer modeling may be needed to explain B84's emission geometry.

\section{HCO$^+$ emission properties \label{sec:hcop}}

\begin{figure*}
\begin{center}
\includegraphics{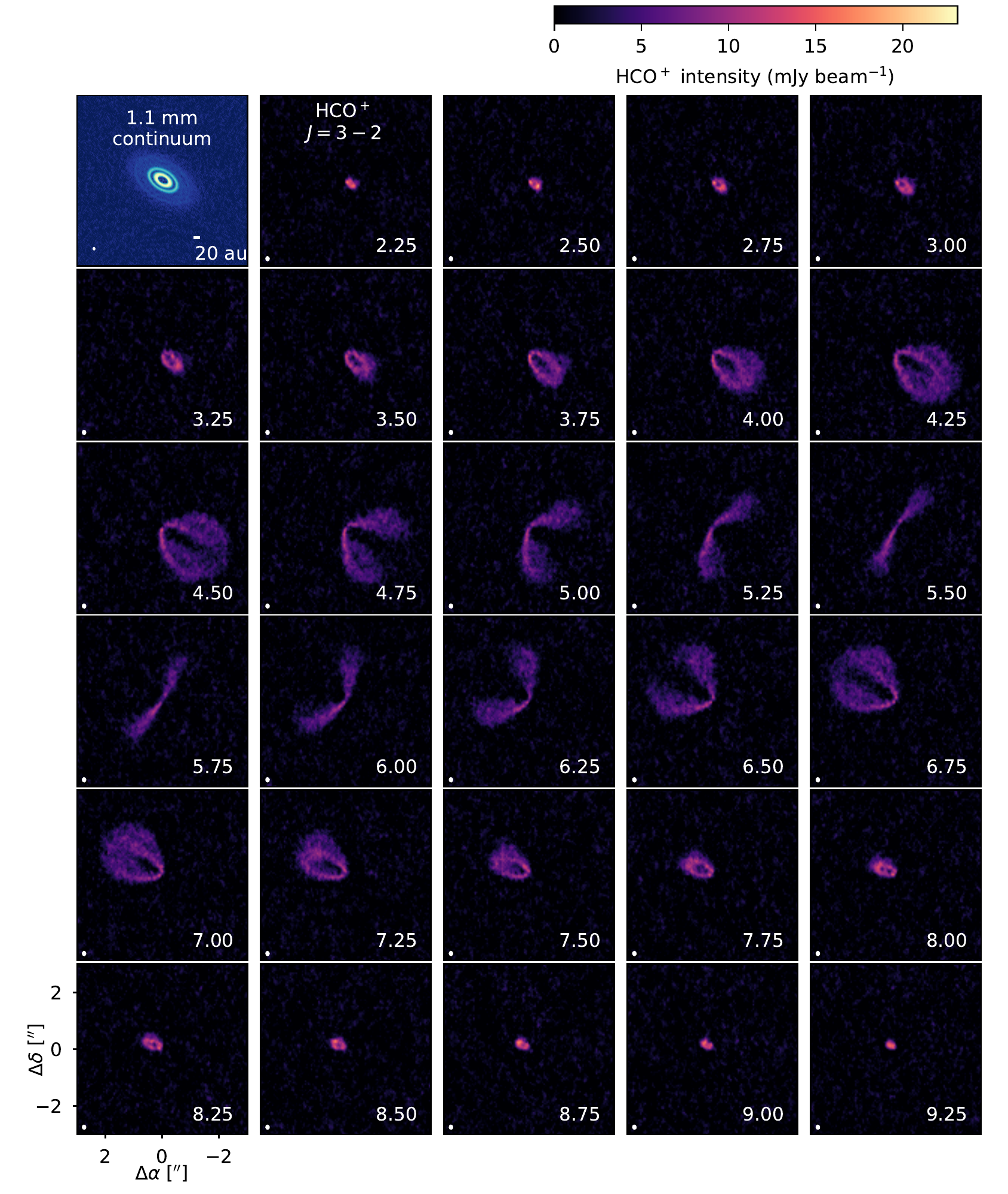}
\end{center}
\caption{Channel maps of the HCO$^{+}$ $J=3-2$ emission, along with the 1.1 mm continuum emission shown in the upper left corner for scale. Synthesized beams and the LSRK velocities (km s$^{-1}$) are shown in the corners of each panel.  \label{fig:HCOpchanmap}}
\end{figure*}

Channel maps of the HCO$^+$ $J=3-2$ emission are shown in Figure \ref{fig:HCOpchanmap}. Because of the favorable viewing angle, one can observe the bright upper surface and dimmer lower surface layers that are characteristic of optically thick line emission in highly flared disks \citep[e.g.,][]{2013ApJ...774...16R,2018AA...609A..47P}. 

\begin{figure}
\begin{center}
\includegraphics{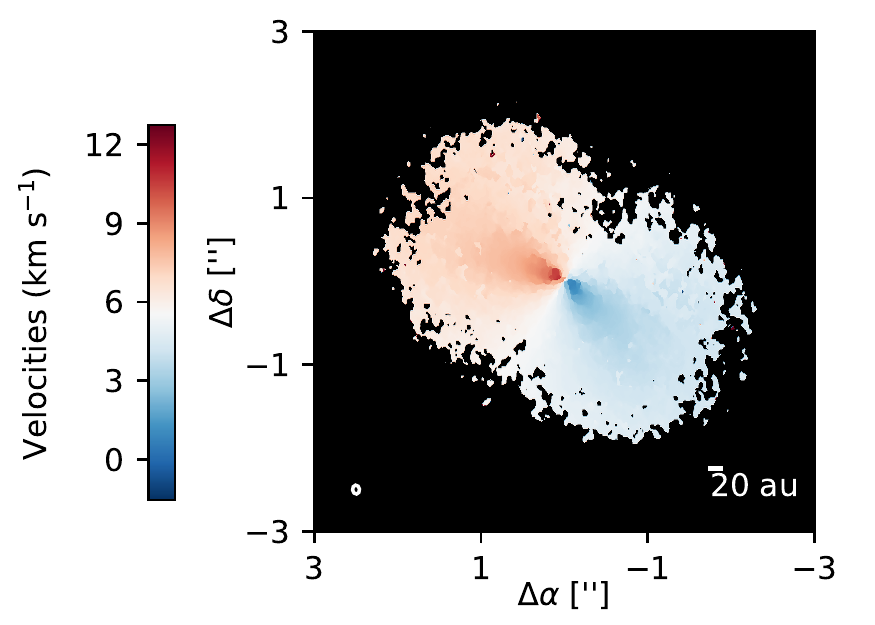}
\end{center}
\caption{A map of the line-of-sight velocities at the emission line peak of each pixel. Pixels where the integrated intensity falls below 2.5 mJy beam$^{-1}$ km s$^{-1}$ are masked. \label{fig:rotationmap}}
\end{figure}

\subsection{The HCO$^+$ emitting height}
We use the Python packages \texttt{bettermoments} and \texttt{eddy} \citep{2019ascl.soft01009T, 2019JOSS....4.1220T} to estimate the height of the HCO$^+$ emission layer. The line-of-sight velocity corresponding to the emission line peak at each pixel is computed using a quadratic fit to the intensities as a function of LSRK velocity (Fig. \ref{fig:rotationmap}). This is functionally similar to an intensity-weighted velocity map (also known as a moment 1 map), but without emission from the lower surface biasing the estimate of the velocities of the brighter upper surface. The resulting map is then downsampled by 10 pixels (0$\farcs$1) on each side such that there is approximately one pixel remaining for each beam-sized patch. The height of the emitting layer as a function of disk radius is parametrized as 

\begin{equation}
z(r) = z_0 \left ( \frac{r}{1''}\right)^\psi.
\end{equation}
Assuming Keplerian kinematics, the quantities needed to compute the line-of-sight velocities for a given disk geometry are $z_0$, $\psi$, the stellar mass $M_\ast$, the systemic velocity $v_\text{LSR}$, the P. A., inclination, and offset from the phase center. The P. A., inclination, and phase offsets are fixed to the values derived from the continuum visibility modeling described in Section \ref{sec:surfbrightness}, since the SNR is higher for the continuum data. Broad uniform priors are adopted for the four free parameters and are listed in Table \ref{tab:velocitypriors}.

The \texttt{eddy} backend uses \texttt{emcee} to fit the observed line-of-sight velocity map with the model Keplerian velocity map. The region of the fit is restricted to radii extending from $0\farcs2$ to $2''$, where the inner radius is set by angular resolution limitations and the outer radius is set by signal-to-noise ratio limitations and to avoid confusion from the dimmer back side of the disk. The posterior probability distributions are sampled with 48 walkers for 3500 steps, with the first 500 steps discarded as burn-in. Convergence is checked by estimating the autocorrelation time for each parameter, which is $\sim$ 40 steps. The posterior medians are listed in Table \ref{tab:velocitypriors}, with error bars calculated from the 16th and 84th percentiles. As noted in \citet{2019AA...625A.118K}, the nominal error bars should be regarded with caution because there may also be systematic uncertainties associated with image-plane fitting and differences between the assumed model structure and true emission behavior. 

\begin{deluxetable}{cccc}
\tablecaption{HCO$^+$ emission height model \label{tab:velocitypriors}}
\tablehead{
\colhead{Parameter}&\colhead{Prior}&\colhead{Results}&\colhead{Units}}
\startdata
$z_0$ &[0, 5]&$0.127\pm0.002$& Arcseconds \\
$\psi$ &[0, 5]&$0.81\pm0.03$& Dimensionless\\
$M_\ast$ &[0.1, 2]&$1.206\pm0.004$& $M_\odot$\\
$v_\text{LSR}$ &[0, 12]&$5.612\pm0.003$& km s$^{-1}$\\
\enddata
\end{deluxetable}

\begin{figure*}
\begin{center}
\includegraphics{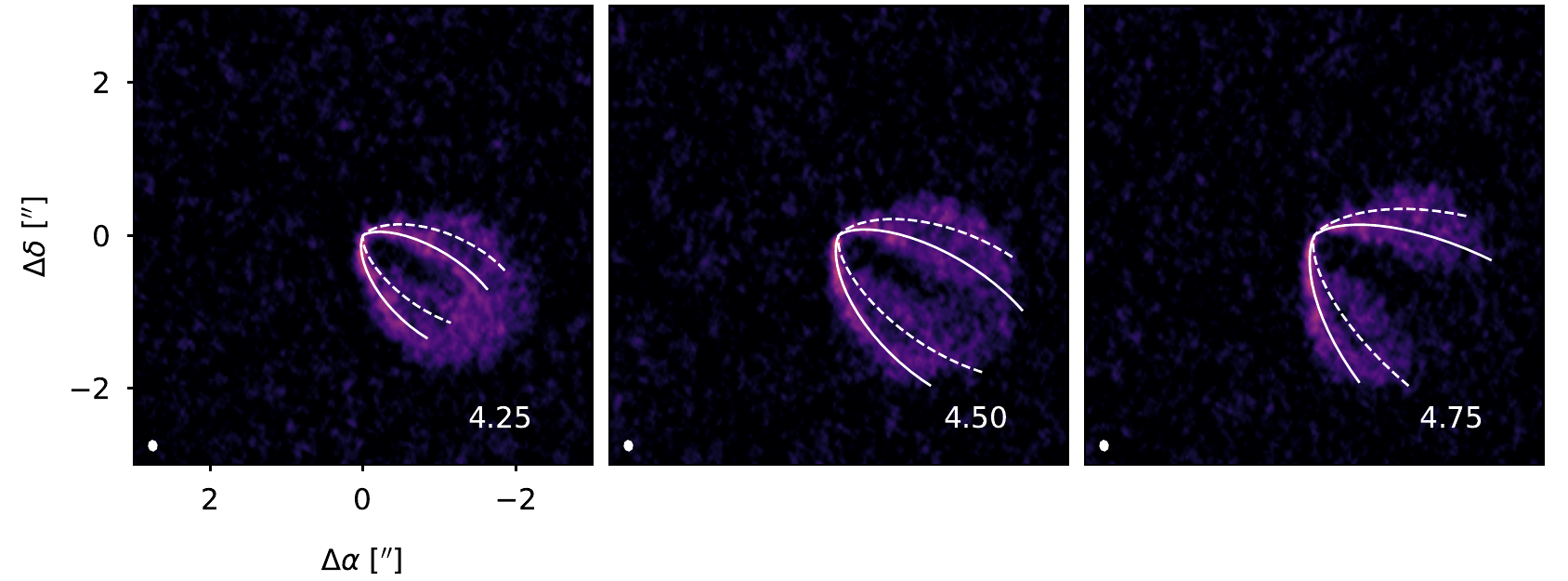}
\end{center}
\caption{Isovelocity contours corresponding to the best-fit model for the HCO$^+$ emitting height are plotted over selected channels of the observed emission. Solid curves denote the front of the disk and dashed curves mark the back of the disk. Synthesized beams are drawn in the lower left corner and LSRK velocities (km s$^{-1}$) are noted in the lower right corner. \label{fig:isovelocities}}
\end{figure*}

The best-fit model adequately reproduces the geometry of the HCO$^+$ emission, as shown in the comparison of the predicted isovelocity contours to channels tracing the front and back sides of the disk in Figure \ref{fig:isovelocities}. The emission geometry demonstrates that the northwest side of the disk is tilted toward the observer, consistent with conclusions drawn from scattered light observations \citep{2003AJ....125.1467S}. While \citet{2008AA...490L..15D} and \citet{2009ApJ...698..131H} identify a prominent discrepancy between the position angles of the continuum and $^{12}$CO emission, no such discrepancy is apparent for HCO$^+$. This may be a consequence of the CO and HCO$^+$ emission coming from different heights in the disk. Interpretation of GM Aur's $^{12}$CO emission has also been complicated by absorption from either intervening cloud or residual envelope material  \citep[e.g.,][]{2009ApJ...698..131H}. No cloud contamination or envelope material is seen in HCO$^+$ emission. With a high critical density of $\sim10^6$ cm$^{-3}$ \citep{2015PASP..127..299S}, the HCO$^+$ $J=3-2$ line should preferentially trace the denser disk material. Based on visual inspection of the channel maps and line-of-sight velocities map, we do not identify clear signatures of a velocity ``twist'' associated with warps and radial inflows \citep[e.g.,][]{2014ApJ...782...62R} or a spectrally and spatially localized perturbation from Keplerian velocities \citep{2018ApJ...860L..13P}. However, since the detection criteria for such features have not been established in detail, our statement on the presence or absence of these features is not intended to be definitive. 

\subsection{Constraints on the HCO$^+$ distribution and temperature}
The isovelocity curves corresponding to the best-fit model are used to generate a Keplerian mask \citep[e.g.,][] {2013ApJ...774...16R, 2017AA...606A.125S} to apply to the image cube before integrating along the spectral axis to produce an integrated intensity map. The integration range, $-1.25$ to $12.75$ km s$^{-1}$, is selected to encompass channels with disk emission above the 3$\sigma$ level. The flux is measured by summing all the unmasked pixels. The flux uncertainty is estimated by generating 50 image cubes consisting only of noise by randomly drawing line-free channels and applying random position shifts, measuring the flux in each cube with the Keplerian mask applied, then taking the standard deviation of these flux measurements. The integrated intensity map and corresponding deprojected, azimuthally averaged radial profile are shown in Figure \ref{fig:momentmaps}. The HCO$^+$ integrated emission peaks at $\sim$ 20 au, which lies well interior to the peak of the continuum emission ring B40. The small central cavity in HCO$^+$ emission is most likely due to a gas surface density reduction, given that much of the dust is cleared in the inner regions of the disk. \citet{2008AA...490L..15D} inferred a similarly-sized hole from modeling low-resolution CO isotopologue emission. The radial profile of the integrated intensity map shows emission decrements near the locations of the continuum rings B40 and B84. These features appear to be largely if not entirely an artifact of continuum subtraction from optically thick line emission \citep[e.g.,][]{2017ApJ...840...60B}. The unresolved inner disk continuum emission, however, is neither extended enough nor bright enough to be responsible for the central HCO$^+$ cavity.

Spatially resolved, optically thick lines are useful for constraining disk gas temperatures because they require a minimal number of assumptions about disk properties \citep[e.g.,][]{2018AA...609A..47P}. Among other things, accurate temperatures are fundamental for measuring disk masses, molecular abundances, and the properties of embedded protoplanets \citep[e.g.,][]{2017AA...605A..69T, 2018ApJ...867L..14V}. The brightness temperature of the optically thick HCO$^+$ emission in GM Aur can be used to estimate the gas temperature at the best-fit emission height given in Table \ref{tab:velocitypriors}. Using the imaging procedure outlined in Section \ref{sec:observations} and the median-stacking procedure described in \citet{2018ApJ...852..122H}, we produce a peak brightness temperature map of HCO$^+$ without continuum subtraction in order to avoid an artificial reduction in the peak line intensities \citep[e.g.,][]{2017ApJ...840...60B, 2018ApJ...853..113W}. The full Planck equation is used to convert peak intensities to brightness temperatures. A deprojected, azimuthally averaged radial profile of the brightness temperature map is produced using the disk coordinates calculated by \texttt{eddy} when fitting the flared HCO$^+$ surface, since the peak intensity at each pixel emerges from well above the midplane. (Unless specified, other deprojections in this paper use the geometrically thin approximation). The brightness temperature map and corresponding radial profile are shown in Figure \ref{fig:momentmaps}. The profile between $R\sim30-350$ au is approximated well with a power law. Using Levenberg-Marquardt minimization to fit a power law $T(r) = T_{100}\left(\frac{r}{100\text{ au}}\right)^{-q}$ to the brightness temperature profile and sampling at radii spaced approximately one synthesized beam apart from $R=30-350$ au, we obtain best-fit results and 1$\sigma$ errors of $T_{100}=26.9\pm3$ and $q=0.43\pm0.01$. The absolute flux calibration uncertainty contributes an additional 10\% uncertainty on $T_{100}$, although it does not change the shape of the curve. 

\begin{figure*}
\begin{center}
\includegraphics{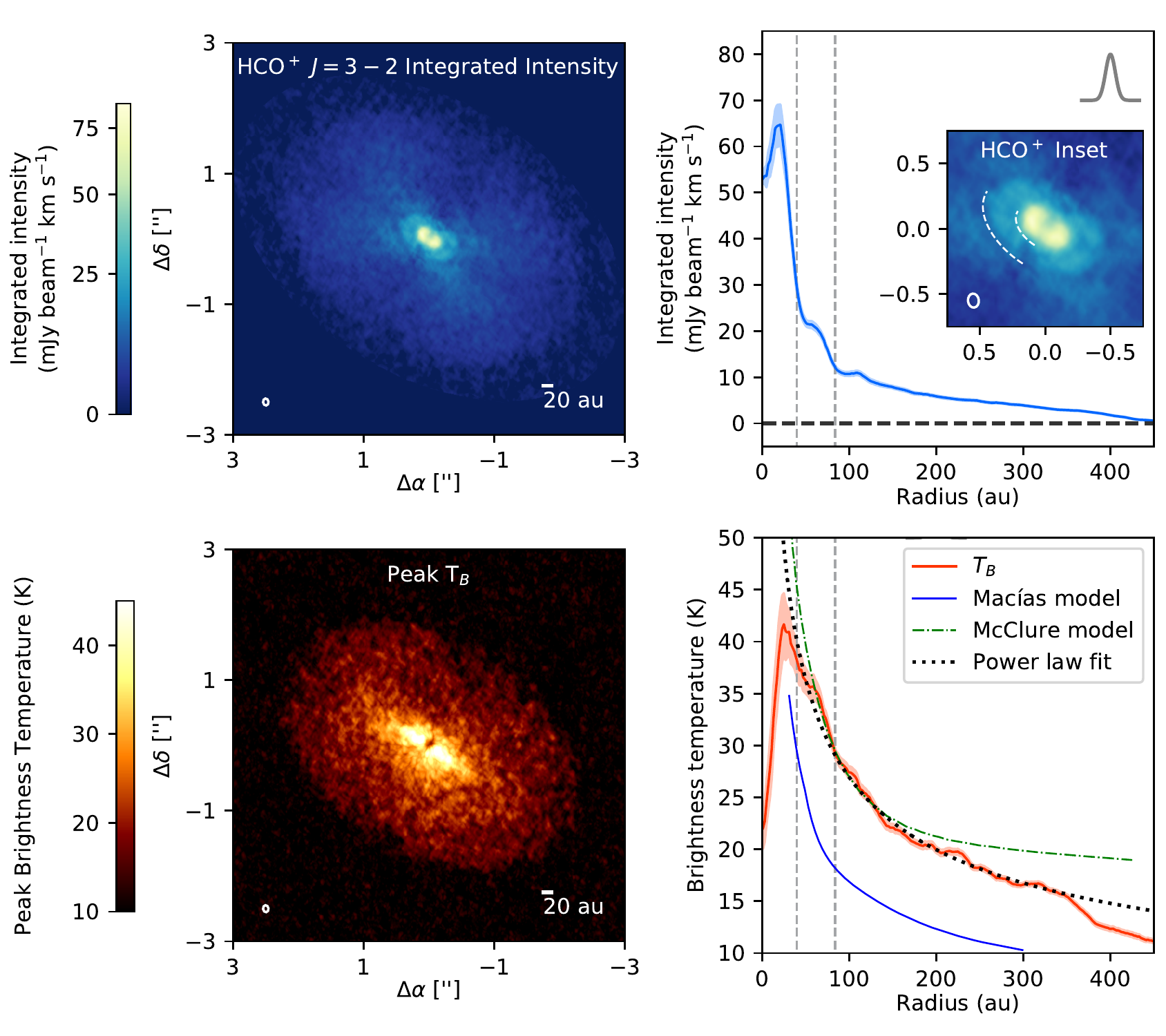}
\end{center}
\caption{\textit{Top left}: Integrated intensity map of HCO$^+$ $J=3-2$ emission in the GM Aur disk. The synthesized beam is drawn in the lower left corner. \textit{Top right}: Deprojected, azimuthally averaged radial profile of the integrated intensity. The vertical dashed lines mark the locations of the continuum rings B40 and B84. The inset panel shows the inner regions of the integrated intensity map, with dashed arcs marking the locations of the emission decrements coinciding with the locations of B40 and B84. The scale of the synthesized beam is shown with the Gaussian in the upper right corner. Shaded ribbons show the 1$\sigma$ scatter at each elliptical bin divided by the square root of the number of beams spanning the bin. \textit{Bottom left}: Similar to top left, but for the brightness temperature. \textit{Bottom right}: Similar to top right, but for the brightness temperature. The dashed black curve shows the power law fit to the observed brightness temperature profile. The solid blue curve shows the \citet{2018ApJ...865...37M} model gas temperature values (which extend from 31 to 300 au) corresponding to the estimated height of the HCO$^+$ emission. The broken green curve similarly shows the \citet{2016ApJ...831..167M} gas temperature model values (which extend from 16 to 428 au). \label{fig:momentmaps}}
\end{figure*}

We can compare the gas temperature estimated from the optically thick HCO$^+$ emitting layer to GM Aur disk temperature models. Using a linear interpolation of the two-dimensional temperature structures from \citet{2016ApJ...831..167M} and \citet{2018ApJ...865...37M}, we plot the temperatures corresponding to the estimated height of HCO$^+$ alongside the brightness temperature profile in Figure \ref{fig:momentmaps}. The \citet{2016ApJ...831..167M} temperatures are a good match to the HCO$^+$ brightness temperature within $\sim100$ au, but increasingly deviate outside 100 au. The \citet{2018ApJ...865...37M} temperatures are $\sim30-40\%$ lower than the HCO$^+$ brightness temperatures, a discrepancy that is too large to be explained by the typically quoted ALMA flux calibration uncertainties ($\sim10\%$).  The key difference between the models is that the \citet{2016ApJ...831..167M} model invokes a much higher depletion of small dust grains in the disk upper layers compared to \citet{2018ApJ...865...37M}. The \citet{2018ApJ...865...37M} model is constrained by the SED and by spatially resolved millimeter and centimeter observations, which are sensitive to the inner disk and the disk midplane rather than the intermediate disk layers traced by HCO$^+$. Meanwhile, the \citet{2016ApJ...831..167M} model is constrained by the SED and HD line emission, which originates from intermediate disk layers. Thus, the  \citet{2016ApJ...831..167M} model has a steep vertical temperature gradient in order to make the intermediate disk layers sufficiently warm to match the HD flux, while the \citet{2018ApJ...865...37M} temperature structure is nearly vertically isothermal from the midplane to $z/r\sim0.2$. (The models are also calculated at slightly different distances, 140 pc for \citet{2016ApJ...831..167M} vs.  160 pc for \citet{2018ApJ...865...37M}, but this does not account for the large difference in temperature structures). The difference between these two models highlights the importance of developing an independent method of inferring temperature. Although a single HCO$^+$ transition cannot be used to determine the shape of the vertical temperature gradient, this exercise illustrates the potential for constraining GM Aur's thermal structure through high-resolution observations of multiple optically thick lines probing different heights.

It should be noted that the temperature differences between the two models are not as large at the midplane. At a given radius, the midplane temperature from \citet{2016ApJ...831..167M} is typically a few degrees \textit{cooler} than that of \citet{2018ApJ...865...37M} (i.e., the opposite of the behavior at the estimated height of the HCO$^+$ layer.) Because continuum emission is in principle more sensitive to the midplane temperature than HD emission, we used the \citet{2018ApJ...865...37M} thermal structure to analyze the dust properties in Section \ref{sec:dustmodels}. However, had we used the \citet{2016ApJ...831..167M} thermal structure instead, we would still conclude that it is ambiguous the extent to which the spectral index profiles signify that the rings are dust traps. The dust optical depth would increase everywhere, but the gaps and the outer disk ($R\gtrsim100$ au) would still be optically thin. Because $\beta$ is only weakly dependent on temperature, the dust grain size constraints are similar for these optically thin regions. With the increased optical depths at the peaks of the emission rings, it remains possible to produce low spectral indices with relatively small $a_\text{max}$ values (e.g., much smaller than 1 mm), but the presence of grains larger than a millimeter is not ruled out. Surface densities remain uncertain due to the possibility that the emission is saturated.

\section{Discussion \label{sec:discussion}}

\subsection{Origin of the emission rings}
As one of the earliest transition disks to be characterized in detail, the GM Aur disk has long been hypothesized to host at least one giant protoplanet \citep[e.g.,][]{1992ApJ...395L.115M, 2003MNRAS.342...79R}. To date, no directly imaged protoplanet candidate has been reported for GM Aur, nor for most other disks. However, the interpretation of gaps in protoplanetary disks as being due to planet-disk interactions has been encouraged by the close match between hydrodynamical simulations and observations \citep[e.g.,][]{2017ApJ...850..201B, 2018ApJ...869L..47Z}, protoplanet detections in the PDS 70 disk cavity \citep{2018AA...617A..44K,2019NatAs.tmp..329H}, and identification of kinematic perturbations in the vicinity of disk gaps \citep[e.g.,][]{2018ApJ...860L..12T,2019NatAs...3.1109P}. 

Using the relationships between gap widths and planet masses derived in \citet{2018ApJ...869L..47Z} from a hydrodynamical parameter space study of non-migrating protoplanets in disks, we can perform a back-of-the-envelope estimate of the mass of a planet that might create GM Aur's D67 gap. The relationships are calculated from models at 1.27 mm, so we use the 1.1 mm map of GM Aur to calculate the gap width because the intensity profiles should not significantly differ over this small wavelength range. We use the observations directly rather than our surface brightness model to compute the width because the \citet{2018ApJ...869L..47Z} relationships are given for intensity maps convolved with Gaussian kernels comparable in size to the GM Aur synthesized beam. As defined in \citet{2018ApJ...869L..47Z}, the normalized width of D67 is $\Delta=0.23$. Based on the SED and millimeter continuum observations, \citet{2018ApJ...865...37M} estimate a gas surface density of $\sim23$ g cm$^{-2}$ and a midplane temperature of $\sim20$ K at a radius of 67 au. Assuming hydrostatic equilibrium and adopting a stellar mass of 1.32 M$_\odot$ \citep{2018ApJ...865..157A}, this would correspond to $h/r\sim0.06$ at 67 au. If the maximum grain size is $\sim1$ mm, a planet with a mass between $\sim$0.1 and 0.4 M$_J$ would be required to open the gap, assuming that the $\alpha$ viscosity parameter ranges between 10$^{-4}$ and 10$^{-2}$. 

Photoevaporation is another mechanism that can clear the inner regions of disks \citep[e.g.,][]{2001MNRAS.328..485C, 2011MNRAS.410..671E, 2012MNRAS.422.1880O}. Traditionally, it has not been regarded as a mechanism likely to create GM Aur's inner cavity, given its large radius and the high accretion rate onto the star \citep[e.g.,][]{2010ApJ...717..441E}. More recently, \citet{2018MNRAS.473L..64E} and \citet{2019MNRAS.490.5596W} have shown that for disks depleted in gas-phase carbon and oxygen, X-ray photoevaporation can open gaps at tens of au, compared to $\sim 1$ au at standard metallicity. The inner disks, and therefore high accretion rates, can be sustained longer when the disk metallicity is decreased.

Annular substructures have also been hypothesized to trace the locations of molecular snowlines because the freezeout of different volatiles is expected to modify the fragmentation and coagulation properties of dust grains \citep[e.g.,][]{2015ApJ...806L...7Z, 2016ApJ...821...82O, 2017ApJ...845...68P}. \citet{2019ApJ...882..160Q} found that the inner and outer edges of N$_2$H$^+$ emission in the GM Aur disk coincided with the continuum rings B40 and B84. The inner and outer boundaries of N$_2$H$^+$ were hypothesized to be set by the CO and N$_2$ snowlines, respectively. An alternative interpretation of the N$_2$H$^+$ observations toward GM Aur could be that ionization is enhanced inside large disk gaps, which was previously invoked by \citet{2019ApJ...871..107F} to explain the rise of DCO$^+$ emission inside one of AS 209's disk gaps. This scenario needs to be tested by source-specific thermochemical models, since the emission geometry of N$_2$H$^+$ is highly sensitive to both ionization and the thermal structure \citep[e.g.,][]{2019ApJ...882..160Q}. While the N$_2$H$^+$ distribution may suggest an association between the dust substructures and snowlines, the appearance of the GM Aur disk in scattered light does not. Models from \citet{2017ApJ...845...68P} indicate that snowline-induced dust substructures should be deeper and wider in near-infrared scattered light images compared to millimeter/sub-millimeter images. However, no clear substructures are observed in the Subaru HiCIAO $H-$band polarized intensity image of GM Aur from \citet{2016ApJ...831L...7O}. Another useful test may be to investigate whether the B84 ring is indeed warped, since snowlines are not expected to induce warps while planets may. 

\subsection{Origin of the central compact emission \label{sec:innerdisk}}
The new ALMA observations reveal an unresolved emission source inside the central cavity of the GM Aur disk. The flux of this feature is not straightforward to define due to the more extended low-lying emission in the cavity at 1.1 mm, but we can compare the two ALMA bands by defining a common area over which the flux is measured. PSF artifacts are a concern for faint emission features, particularly inside a bright emission cavity observed with ALMA's long baselines, but the observations at the two bands are presumably affected similarly because they were observed with the same array configurations. Based on the continuum images used to generate the high-resolution spectral index profiles in Figure \ref{fig:comparebands}, the flux within a diameter of 100 mas is $0.28\pm0.02$ mJy at 1.1 mm and $0.09\pm0.02$ mJy at 2.1 mm. The measurement region is chosen to be slightly larger than the synthesized beam. These fluxes correspond to a spectral index of $1.9\substack{+ 0.5\\-0.4}$. 

While the absolute uncertainties on the spectral index are large, the normalized continuum profiles in the two bands (Figure \ref{fig:comparebands}) indicate that the intensity changes less steeply as a function of frequency at the disk center compared to the emission rings. The presence of larger particles in the inner disk may explain this difference, since large dust grains are expected to drift rapidly toward the star \citep[e.g.,][]{1977ApSS..51..153W}. Given that SED models of the GM Aur disk consistently find that the inner disk is highly dust-depleted \citep[e.g.,][]{1997ApJ...490..368C, 2005ApJ...630L.185C,2010ApJ...717..441E}, it seems less likely that high optical depth can explain the lower spectral index of the inner emission. Nevertheless, if all of the continuum emission is due to thermal dust emission, then SED model results appear to have underestimated the dust content in the cavity. The best-fit SED model of GM Aur from \citet{2018ApJ...865...37M}, which is based on \citet{2010ApJ...717..441E} but adjusted for the new \textit{Gaia} distance, infers the presence of an optically thin inner disk extending from 0.17 to 0.85 au. Their best-fit inner disk dust mass of $\sim4\times10^{-12}$ M$_\odot$ (assuming $a_\text{min}=0.005$ $\mu$m, $a_\text{max}=0.25$ $\mu$m, and $p=3.5$) yields a 2.1 mm flux of $\sim10^{-3}$ $\mu$Jy, which is many orders of magnitude lower than the observed flux. \citet{2013ApJ...775..114M} show that it is difficult to determine from SEDs whether millimeter-sized grains are present in the inner disk due to their minor contribution to the near-infrared excess. Thus, GM Aur's SED does not necessarily rule out the presence of a millimeter (or larger) grain population that is responsible for the emission detected by ALMA. 

It has been an enduring puzzle how disks such as GM Aur maintain high accretion rates inside strongly depleted dust cavities \citep[e.g.,][]{2014ApJ...782...62R, 2014prpl.conf..497E}.  Thus, it is interesting to consider how much material would have to be contained in the inner disk to sustain GM Aur's accretion rates, which have been measured to range from $3.9\times10^{-9}$ $M_\odot$ yr$^{-1}$ to $1.96\times 10^{-8}$ $M_\odot$ yr$^{-1}$ \citep{2015ApJ...805..149I,2019ApJ...874..129R}. Assuming a gas-to-dust ratio of 100, maintaining a representative accretion rate of $\sim10^{-8}$ $M_\odot$ yr$^{-1}$ for $10^5$ years would require a dust mass of 10$^{-5}$ $M_\odot$ in the inner disk if it is not replenished by the outer disk. This would represent $0.5-4\%$ of GM Aur's estimated total dust mass \citep{2016ApJ...831..167M}. Using our power-law fit to the temperature structure from \citet{2018ApJ...865...37M}, we estimate that an inner disk with a radius of 2 au should have $\tau_\text{abs}=\Sigma_d \kappa_\text{abs}/\cos{i} \sim0.1$ to reproduce the measured inner disk flux at 1.1 mm. A dust mass of 10$^{-5}$ $M_\odot$ spread uniformly over this disk would result in $\tau\gg1$ at millimeter wavelengths if we use the DSHARP opacities and assume $a_\text{max}=1$ mm and $p=3.5$. However, above $a_\text{max}\sim1$ mm, opacities drop rapidly. Values of $p$ at or shallower than 2.5 and $a_\text{max}$ of about 1 meter can yield sufficiently low emission. Invoking solids this large, though, is challenging insofar as they are expected to drift into the star within hundreds of years \citep[e.g.][]{1977ApSS..51..153W}. Alternatively, the inner disk might be able to sustain GM Aur's accretion without requiring extremely large solids if the gas-to-dust ratio is significantly enhanced. 

If the dust emission at the disk center is optically thin, contributions from non-dust emission can help explain the low spectral index. Using VLA centimeter observations of GM Aur, \citet{2016ApJ...829....1M} estimated that free-free emission from ionized gas contributed 76 $\mu$Jy to the flux measured at 3.0 cm. This emission was attributed to a combination of an ionized radio jet and a photoevaporative wind in the inner disk. Extrapolating the amount of free-free emission expected at millimeter wavelengths from existing data is not straightforward because the spectral index changes as the free-free emission becomes optically thin, which typically occurs between 1 and 10 cm \citep[e.g.,][]{2018ApJ...860...77E}. Furthermore, it is unknown whether the free-free emission measured with the VLA is co-spatial with  the inner disk emission measured with ALMA because the synthesized beam for the 3 cm observations is $10\times$ larger. To make a conservative estimate of whether free-free emission could be detectable at ALMA wavelengths, we assume that it becomes optically thin at 3.0 cm. Since optically thin free-free emission scales as $F_{\nu, \text{ff}}\propto \nu^{-0.1}$ \citep[e.g.,][]{2008ApJ...683..304E}, the flux due to free-free emission is expected to be $\sim58$ $\mu$Jy at 2.1 mm and $\sim55$ $\mu$Jy at 1.1 mm. This represents about two-thirds and one-fifth of the inner disk flux measured at 2.1 and 1.1 mm, respectively. The spectral index of the inner disk is within the range computed for optically thick winds \citep[e.g.,][]{1986ApJ...304..713R}, but it is unlikely that the inner disk millimeter emission of GM Aur originates from optically thick free-free emission. Using the $\alpha_\text{free-free}=0.75$ from centimeter observations of GM Aur in \citet{2016ApJ...829....1M}, a direct extrapolation results in a millimeter wavelength flux that is several times higher than the measured inner disk flux. Magnetic reconnection in the stellar corona is also hypothesized to generate significant emission at millimeter wavelengths \citep[e.g.,][]{2006AA...453..959M, 2008AA...492L..21S}, but this mechanism is usually associated with variability on timescales of a few hours, whereas the GM Aur fluxes are consistent between execution blocks. Ultimately, to disentangle the contributions of dust emission from other sources of emission in the inner disk, high angular resolution observations at other frequencies are necessary. 

\subsection{Comparison with other disks}
\subsubsection{Ring properties}
Annular gaps and rings, which appear to be the most common type of dust substructure, have now been detected in the millimeter continuum of dozens of disks  \citep[e.g.,][and references therein]{2015ApJ...808L...3A, 2018ApJ...869...17L,2018ApJ...869L..41A, 2018ApJ...869L..42H}. While the widths, amplitudes, and locations of these structures are quite varied, the GM Aur millimeter continuum shares some striking characteristics with the DM Tau and AA Tau disks. All three of these disks exhibit strong dust depletion at the center of the disk, one or more narrow, high-contrast rings (i.e., order of magnitude contrast or more) inside a radius of $\sim100$ au, and faint, extended emission beyond 100 au \citep{2017ApJ...840...23L, 2018ApJ...868L...5K}. Qualitatively, a steep drop in surface brightness accompanied by faint extended emission is predicted to be a signature of viscous spreading \citep{2019MNRAS.486.4829R}. Quantitatively, though, the faint outer emission of these three disks is still brighter than the outer millimeter continuum halo computed for models of viscous spreading in \citet{2019MNRAS.486.4829R}. 

Surface brightness models of disks have often approximated disk substructures as Gaussian rings \citep[e.g.,][]{2016PhRvL.117y1101I, 2017ApJ...840...23L,2018ApJ...869L..48G}. However, as the resolution and sensitivity of observations improve, the shapes of some substructures can clearly be distinguished from Gaussians. For both B40 and B84 in the GM Aur disk, the intensity profile is steeper on the sides facing the star. Similar behavior has been noted for several other disks, including DM Tau \citep{2018ApJ...868L...5K}, T Cha \citep{2018MNRAS.475L..62H}, and SR 24S \citep{2019ApJ...878...16P}. In some of these cases, such as for B40 in the GM Aur disk and for the SR 24S disk \citep{2019ApJ...878...16P}, there also appear to be unresolved substructures within a ring structure, which contributes to the radial asymmetry. While recent analyses of the origins of annular substructures have primarily focused on their radial locations and amplitudes \citep[e.g.,][]{2018ApJ...869...17L, 2018ApJ...869L..42H}, well-resolved ring emission profiles hold promise for clarifying formation mechanisms.  \citet{2018ApJ...859...32P} and \citet{2018ApJ...869L..46D} use 1-D disk dust evolution models to show that a planet can induce a ring structure outside its orbit, with the side of the ring facing the planet being steeper. Thus, GM Aur's millimeter emission is qualitatively consistent with expectations for planet-induced substructures. That being said, the millimeter continuum emission of the PDS 70 disk, which hosts two directly imaged protoplanets, exhibits behavior opposite to that of GM Aur \citep{2019AA...625A.118K, 2019NatAs...3..749H}. This could be due to their age difference, since these dust evolution models also predict that planet-induced rings will become narrower and less asymmetric over time due to the depletion of dust in the outer disk. PDS 70, which is estimated to be $5.4\pm1.0$ Myr old \citep{2018AA...617L...2M}, is nominally a few Myr older than GM Aur. To test these dust evolution models in more detail, it may be instructive to examine whether there are systematic changes in ring emission profiles with disk age. 
\subsubsection{Central cavity properties}

The GM Aur disk's compact central emission component appears to be typical of disks with central cavities that have been imaged at high resolution. Other sources with a similar feature detected at millimeter/sub-millimeter wavelengths include TW Hya \citep{2016ApJ...820L..40A}, AB Aur \citep{2017ApJ...840...32T}, V1247 Ori \citep{2017ApJ...848L..11K}, MWC 758 \citep{2018ApJ...860..124D}, T Cha \citep{2018MNRAS.475L..62H}, DM Tau \citep{2018ApJ...868L...5K}, SAO 206462 \citep{2018AA...619A.161C}, HD 143006 \citep{2018ApJ...869L..50P}, HD 100546 \citep{2019ApJ...871...48P}, PDS 70 \citep{2019AA...625A.118K}, HD 169142 \citep{2019AJ....158...15P}, and SR 24S \citep{2019ApJ...878...16P}. In a couple systems, such as DM Tau and HD 143006, the central emission originates from a compact but resolved dust ring. In most of these systems, however, the central emission appears to be unresolved/marginally resolved, indicating a maximum extent of a few au. In these cases, it is more ambiguous whether the central feature is due to dust emission or some other source, such as free-free emission \citep[e.g.,][]{2018ApJ...860..124D}. 

While HCO$^+$ is not a straightforward tracer of the disk gas distribution, its bright emission interior to B40 of the GM Aur disk qualitatively indicates that the molecular gas cavity must have a smaller radius than the millimeter dust cavity. This feature is consistent with the finding from \citet{2016ApJ...829...65H} that sub-micron dust grains are present in the disk down to a radius of at least 24 au (compared to a millimeter continuum peak at 40 au), since gas and small dust grains should be well-coupled. Other studies of sources traditionally classified as large-cavity transition disks also typically find that cavities in molecular emission and scattered light are smaller in radius than the millimeter dust cavity \citep[e.g.,][]{2016AA...585A..58V, 2019AA...624A...7V}. This behavior has been thought to arise from planets inside the cavities creating pressure maxima that trap large dust grains while allowing small dust grains and gas to pass through \citep[e.g.,][]{2013AA...560A.111D}. 

Other than the similarities in the inner disk, high resolution observations have shown that ``transitional'' disks are quite heterogeneous. They span a range of spectral types (M through A), appear in several different star-forming regions, and exhibit diverse emission features in the outer disk, including spiral arms, crescent-like asymmetries, and annular gaps and rings.

\subsubsection{Spectral index behavior}
Prior to the discovery of complex disk structures, low spectral indices were usually attributed to the presence of large dust grains because the disks appeared to be optically thin and the measured disk sizes showed a strong wavelength dependence \citep[e.g.,][]{2001ApJ...554.1087T, 2010AA...512A..15R, 2012ApJ...760L..17P}. This interpretation has become less certain in light of the discovery that many disks have dust concentrated into narrow rings, which translates into relatively large local optical depths but low apparent optical depths at coarse angular resolution due to the small filling factor \citep[e.g.,][]{2017ApJ...845...44T}. Separately, grain size estimates on the order of $a_\text{max}=100$ $\mu$m from polarization studies have also challenged the interpretation of low spectral indices as evidence of grain growth to millimeter/centimeter sizes \citep[e.g.,][]{2016ApJ...820...54K}. 

Moderate to high resolution (i.e., 20 au or better) millimeter wavelength spectral index measurements have been published for only a handful of protoplanetary disks: HL Tau \citep{2015ApJ...808L...3A, 2019ApJ...883...71C}, TW Hya \citep{2016ApJ...829L..35T,2018ApJ...852..122H}, SAO 206462 \citep{2018AA...619A.161C}, HD 163296 \citep{2019MNRAS.482L..29D}, HD 169142 \citep{2019ApJ...881..159M}, SR 24S \citep{2019ApJ...878...16P}, and GM Aur (this work). For most of these sources, the spectral index rises inside continuum gaps and decreases at the continuum rings. Interpretations of this behavior have varied\textemdash \citet{2018ApJ...852..122H}, \citet{2019ApJ...878...16P}, and \citet{2019ApJ...883...71C} suggest that optical depth variations are a significant contributor to the spectral index variations in the individual disks they analyze, while \citet{2016ApJ...829L..35T} and \citet{2019ApJ...881..159M} make the case for spectral index variations being due largely to dust trapping or dust filtration preferentially segregating large grains inside ring structures \citep[e.g.,][]{2006MNRAS.373.1619R}.  

These differences in interpretation may be due at least in part to different methodologies (e.g., choice of dust opacities, thermal structure calculations, accounting for scattering, etc.), but could also be due to bona fide structural variations between disks. \citet{2019ApJ...883...71C} argue that HL Tau is optically thick at millimeter wavelengths not only at the emission ring peaks, but also in the gaps. This conclusion differs from the analyses of the other aforementioned disks, which find that the emission gaps are optically thin even if the optical depths of the rings are ambiguous. However, given that HL Tau is a young, ``flat spectrum'' object surrounded by significant envelope material \citep[e.g.,][]{2007ApJS..169..328R}, it would not be surprising if its dust properties differ from the more evolved Class II disks. 

For the few disks observed at very high resolution ($\sim$ a few au), a subtle difference in spectral index behavior appears. The spectral index radial profiles of the TW Hya and HL Tau disks feature sections that are ``flattened'' at a value of $\alpha\sim2$ over a span of several au or more, whereas the GM Aur disk's spectral index profile changes sharply around its local minima. While this distinction may simply be a consequence of different underlying surface density distributions, another interpretation is that the ``flattening'' of TW Hya and HL Tau's spectral index radial profile at low values of $\alpha$ might be due to the emission saturating (in other words, the dust is optically thick at these radii). This is linked to the idea from \citet{2018ApJ...869L..46D} that optically thick and thin rings might be distinguished by their emission profiles, since optically thick substructures should saturate around their surface density peaks and produce ``flat-topped'' rings. High-resolution, high-sensitivity observations at longer, more optically thin wavelengths (e.g., with the planned ngVLA) will be valuable for clarifying the origins of the spatially varying spectral indices in disks.

\section{Summary\label{sec:summary}}
We present the highest resolution millimeter continuum observations to date of the GM Aur disk in conjunction with HCO$^+$ $J=3-2$ observations. The multi-frequency continuum observations are used to probe the dust properties, while the HCO$^+$ observations are used to examine gas properties. Our main results are as follows: 
\begin{enumerate}
\item The GM Aur dust disk is highly structured. The 1.1 mm continuum features rings at $\sim40$, 84, and 168 au, as well as faint unresolved emission inside the central cavity and faint extended emission in the outer disk. The 2.1 mm continuum is similar, but an additional shoulder is observed on the B40 ring, and the B168 ring is not clearly detected. 
\item The radial spectral index profile features local minima near continuum rings and local maxima near continuum gaps, similar to behavior seen in the handful of disks that have been observed at multiple wavelengths at high resolution. We model the visibilities to extract the surface brightness profiles and use the dust temperature model from \citet{2018ApJ...865...37M} to derive the dust optical depth. We find that the gaps and the diffuse outer emission ($R\gtrsim100$ au), including the B168 emission ring, are optically thin. The optical depths at the peaks of the B40 and B84 emission rings are high enough that scattering should be explicitly taken into account when relating the measured intensities to the dust grain sizes. From the current data, it is ambiguous whether the radial spectral index variations trace the trapping of large grains in B40 and B84, or whether high optical depth alone is responsible for the low spectral indices. However, the different spectral indices at the peaks of B40 and B84, as well as spectral index variations in the optically thin outer disk, indicate that the GM Aur millimeter continuum emission is not compatible with a radially uniform dust population. 
\item A comparison of the best-fit continuum surface brightness model to the 1.1 mm observations indicates that the inner edge of B84 appears to be at a different inclination than the outer edge. We use RADMC-3D radiative transfer models to demonstrate that this emission geometry might be a consequence of vertical structure or a mild warp. 
\item Like other transition disks that have been imaged at high resolution, GM Aur features compact emission inside the central cavity. While the presence of an optically thin dust disk has previously been inferred from SED modeling, it is insufficient to explain the millimeter emission detected. We posit that GM Aur has a population of large dust grains (of order millimeter size or more) in the inner disk, with contributions from free-free emission. 
\item HCO$^+$ $J=3-2$ emission in the GM Aur disk is bright, extended, and flared. The HCO$^+$ emission cavity, and by extension the gas cavity, have smaller radii than the millimeter dust cavity. The HCO$^+$ brightness temperatures indicate that the disk layer it emerges from is fairly warm ($T\sim27$ K at $R=100$ au). We advocate for using similar high-resolution imaging of other optically thick lines to constrain the thermal structure of the GM Aur disk. 
\end{enumerate}

While high resolution ($<10$ au) ALMA observations of disks now number in the dozens, the GM Aur disk joins only a handful of sources that have been mapped in high resolution at more than one frequency. These observations collectively demonstrate that the radial spectral index profile is closely tied to disk substructures, and thus observations that do not resolve disk substructures can yield inaccurate estimates of disk optical depths and grain sizes. In particular, high resolution observations have shown that local disk optical depths are higher than estimates from low-resolution observations, so the grain sizes in regions associated with low spectral index values ($\alpha<2.5$) may be smaller than previously inferred. Access to high-resolution observations at more than one frequency also provides a more comprehensive picture of substructures present in the disk and places constraints on contributions from non-dust emission. Ultimately, to improve constraints on the growth and transport of solids in disks, it is crucial to upgrade the Very Large Array and/or ALMA to enable similarly high-resolution, high-sensitivity imaging of disks at wavelengths where the dust is expected to be optically thin (i.e., 7 mm or longer). 

\acknowledgments
We thank the referee for comments improving this paper. We also thank Meredith Hughes and Xuening Bai for helpful discussions and the NRAO staff for their advice on data calibration and reduction. This paper makes use of ALMA data\\
\dataset[ADS/JAO.ALMA\#2017.1.01151.S]{https://almascience.nrao.edu/aq/?project\_code=2017.1.01151.S} and\\ \dataset[ADS/JAO.ALMA\#2018.1.01230.S]{https://almascience.nrao.edu/aq/?project\_code=2018.1.01230.S}. ALMA is a partnership of ESO (representing its member states), NSF (USA) and NINS (Japan), together with NRC (Canada) and NSC and ASIAA (Taiwan), in cooperation with the Republic of Chile. The Joint ALMA Observatory is operated by ESO, AUI/NRAO and NAOJ. The National Radio Astronomy Observatory is a facility of the National Science Foundation operated under cooperative agreement by Associated Universities, Inc. The computations in this paper were run in part on the Odyssey cluster supported by the FAS Division of Science, Research Computing Group at Harvard University and on the Smithsonian Institution High Performance Cluster (SI/HPC). This research has made use of NASA's Astrophysics Data System. J. H. acknowledges support from the National Science Foundation Graduate Research Fellowship under Grant No. DGE-1144152. S. A. and J. H. acknowledge funding support from the National Aeronautics and Space Administration under Grant No. 17-XRP17\_2-0012 issued through the Exoplanets Research Program. J. M. C. acknowledges support from the National Aeronautics and Space Administration under Grant No. 15XRP15\_20140 issued through the Exoplanets Research Program. L. P. acknowledges support from CONICYT project Basal AFB-170002 and from FONDECYT Iniciaci\'on project \#11181068. Z. Z. acknowledges support from the National Science Foundation under CAREER Grant Number AST-1753168. A.I. acknowledges support from the National Science Foundation under grant No. AST-1715719. 

\facilities{ALMA}

\software{ \texttt{analysisUtils} (\url{https://casaguides.nrao.edu/index.php/Analysis_Utilities}), \texttt{AstroPy} \citep{2013AA...558A..33A}, \texttt{bettermoments} \citep{2019ascl.soft01009T}, \texttt{CASA} \citep{2007ASPC..376..127M}, \texttt{dsharp\_opac} \citep{2018ApJ...869L..45B}, \texttt{eddy} \citep{2019JOSS....4.1220T}, \texttt{emcee} \citep{2013PASP..125..306F}, \texttt{matplotlib} \citep{Hunter:2007}, \texttt{scikit-image} \citep{scikit-image}, \texttt{SciPy} \citep{scipy}}

\appendix
\section{Estimating the emergent intensity}\label{sec:scatteringformula}

Because the dust traced by millimeter emission is presumably confined to a thin midplane layer, we compute intensities using a 1-D vertically isothermal slab approximation. The key formulae are summarized in this appendix. We employ the approach developed in \citet{2019ApJ...883...71C} and \citet{2019ApJ...876....7S} based on the approximation by \citet{1993Icar..106...20M}. Because the intensity formula is derived assuming that scattering is isotropic, whereas the scattering in a disk is likely anisotropic, the scattering opacity $\kappa_\nu^\text{sca}$ is replaced with an effective scattering opacity $\kappa_\nu^\text{sca,eff} = \kappa_\nu^\text{sca}(1-g_\nu)$, where $g_\nu$ is the forward-scattering parameter. Then $\Delta \tau_\nu =(\kappa_\nu^\text{abs}+\kappa_\nu^\text{sca,eff})\Sigma_d$, where $\Sigma_d$ is the disk dust surface density. The effective albedo is $\omega_\nu^\text{eff}=\sfrac{\kappa_\nu^\text{sca,eff}}{(\kappa_\nu^\text{abs}+\kappa_\nu^\text{sca,eff})}$. The viewing geometry is accounted for by $\mu=\cos i$. The emergent intensity is 

\begin{equation}
I_\nu^\text{out}\approx B_\nu(T_d) \left(1-e^{-\frac{\Delta \tau_\nu}{\mu} }+\omega_\nu^\text{eff} F(\Delta\tau_\nu, \omega_\nu^\text{eff}, \mu)\right), 
\end{equation}

where 

\begin{equation}
F(\Delta\tau_\nu, \omega_\nu^\text{eff}, \mu) = \frac{1}{e^{-\sqrt{3}\epsilon_\nu^\text{eff}\Delta\tau_\nu}(\epsilon_\nu^\text{eff}-1)-(\epsilon_\nu^\text{eff}+1)}\times \left( \frac{1-e^{-(\sqrt{3}\epsilon_\nu^\text{eff}+1/\mu)\Delta\tau_\nu}}{\sqrt{3}\epsilon_\nu^\text{eff}\mu+1} + \frac{e^{-\Delta\tau_\nu/\mu}-e^{-\sqrt{3}\epsilon_\nu^\text{eff}\Delta\tau_\nu}}{\sqrt{3}\epsilon_\nu^\text{eff}\mu-1}\right)
\end{equation}

and 

\begin{equation}
\epsilon_\nu^\text{eff} = \sqrt{1-\omega_\nu^\text{eff}}.
\end{equation}

The opacity values $\kappa_\nu^\text{abs}$ and $\kappa_\nu^\text{sca,eff}$ depend on the dust power-law distribution parameters $p$ and $a_\text{max}$. Thus at a given frequency, $I_\nu^\text{out}$ is determined by 5 parameters: $p$, $a_\text{max}$, $T_d$, $\Sigma_d$, and $\mu$.

\bibliographystyle{aasjournal}

\end{document}